\documentclass[aps,prd,twocolumn,nofootinbib,superscriptaddress,preprintnumbers,floatfix]{revtex4-1}
\usepackage{amsfonts,amsmath,amssymb,amsthm,braket,nicefrac}
\usepackage{booktabs,array}
\usepackage[utf8]{inputenc}
\usepackage{float,graphicx,hhline}
\usepackage{algorithm,algpseudocode}
\usepackage{appendix}
\usepackage[dvipsnames]{xcolor}
\usepackage{mathtools}
\usepackage{multirow}
\usepackage{tikz-cd}
\usepackage{slashed}
\usepackage{soul}
\usepackage{pifont}
\usepackage{tikzducks}
\usepackage[caption=false]{subfig}
\usepackage[export]{adjustbox}
\usepackage{enumitem} 
\usepackage{bbm}
\usepackage{dsfont}
\usepackage{hyphenat}
\AtBeginDocument{
\heavyrulewidth=.08em
\lightrulewidth=.05em
\cmidrulewidth=.03em
\belowrulesep=.65ex
\belowbottomsep=0pt
\aboverulesep=.4ex
\abovetopsep=0pt
\cmidrulesep=\doublerulesep
\cmidrulekern=.5em
\defaultaddspace=.5em
}

\usepackage{hyperref}
\hypersetup{
    pdftitle={},
    colorlinks=true,     
    linkcolor=blue,      
    citecolor=blue,      
    filecolor=blue,      
    urlcolor=blue        
}
\usepackage{cleveref}
\Crefname{equation}{Eq.}{Eqs.}

\DeclareMathAlphabet{\mathbbold}{U}{bbold}{m}{n}

\DeclareMathOperator{\Tr}{Tr}

\begin{document}
\title{Gauge-equivariant flow models for sampling in lattice field theories with pseudofermions}

\newcommand{\getMITAffiliation}{\affiliation{Center for Theoretical Physics, Massachusetts Institute of Technology, Cambridge, MA 02139, USA}}
\newcommand{\getNYUAffiliation}{\affiliation{Center for Cosmology and Particle Physics, New York University, New York, NY 10003, USA}}
\newcommand{\getDMAffiliation}{\affiliation{DeepMind, London, UK}}
\newcommand{\getIAIFIAffiliation}{\affiliation{The NSF AI Institute for Artificial Intelligence and Fundamental Interactions}}

\author{Ryan~Abbott}
\getMITAffiliation
\getIAIFIAffiliation
\author{Michael~S.~Albergo}
\getNYUAffiliation
\author{Denis~Boyda}
\affiliation{Argonne Leadership Computing Facility, Argonne National Laboratory, Lemont IL-60439, USA}
\getMITAffiliation
\getIAIFIAffiliation
\author{Kyle~Cranmer}
\getNYUAffiliation
\author{Daniel~C.~Hackett}
\getMITAffiliation
\getIAIFIAffiliation
\author{Gurtej~Kanwar}
\affiliation{Albert Einstein Center, Institute for Theoretical Physics, University of Bern, 3012 Bern, Switzerland}
\getMITAffiliation
\getIAIFIAffiliation
\author{S\'{e}bastien~Racani\`{e}re}
\getDMAffiliation
\author{Danilo~J.~Rezende}
\getDMAffiliation
\author{Fernando~Romero-L\'opez}
\getMITAffiliation
\getIAIFIAffiliation
\author{Phiala~E.~Shanahan}
\getMITAffiliation
\getIAIFIAffiliation
\author{Betsy~Tian}
\getMITAffiliation
\author{Julian~M.~Urban}
\affiliation{Institut f\"ur Theoretische Physik, Universit\"at Heidelberg, Philosophenweg 16, 69120 Heidelberg, Germany}

\preprint{MIT-CTP/5446, INT-PUB-22-017}

\begin{abstract}
This work presents gauge-equivariant architectures for flow-based sampling in fermionic lattice field theories using pseudofermions as stochastic estimators for the fermionic determinant. This is the default approach in state-of-the-art lattice field theory calculations, making this development critical to the practical application of flow models to theories such as QCD.
Methods by which flow-based sampling approaches can be improved via standard techniques such as even/odd preconditioning and the Hasenbusch factorization are also outlined. Numerical demonstrations in two-dimensional U(1) and SU(3) gauge theories with $N_f=2$ flavors of fermions are provided.
\end{abstract}
\maketitle

\section{Introduction}

Lattice quantum field theory (LQFT), particularly lattice quantum chromodynamics, has become an ubiquitous tool in high-energy and nuclear theory~\cite{Detmold:2019ghl,USQCD:2019hyg,Kronfeld:2019nfb,Boyle:2022uba}. Given the extraordinary computational cost of state-of-the-art LQFT studies, advances in the form of more efficient algorithms are of great value~\cite{Joo:2019byq,Boyda:2022nmh}. Recently, significant efforts have been made to design novel algorithms incorporating machine-learned components to accelerate the first stage of LQFT calculations, which involves sampling contributions to the high-dimensional discretized path integral~\cite{Wang2017,Huang:2017,song2017nice,Tanaka:2017niz,levy2018generalizing,Pawlowski:2018qxs,Cossu:2018pxj,Wu:2019,Bachtis:2020dmf,Nagai:2020jar,Tomiya:2021ywc,Bachtis:2021eww,Wu:2021tfb}. 
In particular, bespoke generative flow models~\cite{rezende2016variational,dinh2017density,JMLR:v22:19-1028} tailored to the sampling of LQFT field configurations have been developed and applied in a variety of ways~\cite{LiWang2018NNRG,Albergo:2019eim,Kanwar:2020xzo,Boyda:2020hsi,Hackett:2021idh,Albergo:2021bna,Albergo:2021vyo,Albergo:2022qfi,Nicoli:2020evf,Nicoli:2020njz,Foreman:2021ixr,Foreman:2021ljl,Foreman:2021rhs,DelDebbio:2021qwf,Gabrie:2021tlu,deHaan:2021erb,Lawrence:2021izu,Jin:2022bgq,Pawlowski:2022rdn,Finkenrath:2022ogg,Gerdes:2022eve,Singha:2022lpi,Matthews:2022sds,Caselle:2022acb}, and have been found to address key sampling challenges such as critical slowing-down and topological freezing in some two-dimensional (2D) theories~\cite{Albergo:2019eim,Kanwar:2020xzo,Boyda:2020hsi,Hackett:2021idh,Albergo:2021bna,Albergo:2022qfi}.

In order to use flow-based sampling for LQFTs with fermions, such as QCD, suitable flow architectures to treat both gauge and fermionic degrees of freedom are needed. Several pieces of this puzzle are already in place. First, gauge-equivariant architectures for gauge fields have been developed and applied for 2D Abelian and non-Abelian theories~\cite{favoni2020lattice, Kanwar:2020xzo,Boyda:2020hsi}. Second, different methods to incorporate fermions in flow architectures have been discussed in Ref.~\cite{Albergo:2021bna}, with numerical demonstrations provided in the context of the 2D Yukawa theory. Among the various proposals, integrating out the fermion fields and directly evaluating the resulting fermion determinant is the most straightforward approach; in this case, the target probability distribution is given only in terms of gauge variables. This procedure has found success in proof-of-principle applications to the Schwinger model at criticality~\cite{Albergo:2022qfi}.
However, this approach is not scalable. Specifically, computing the probability density after the fermionic integration via direct methods is not feasible for at-scale studies of theories such as QCD, as such methods scale cubically with the spacetime volume. The usual approach to this challenge is to introduce auxiliary degrees of freedom, named pseudofermions, which function as stochastic determinant estimators for which the cost of evaluation scales more favorably with the lattice volume. It seems natural to follow a similar approach in flow-based sampling. This was considered in Ref.~\cite{Albergo:2021bna} for Yukawa theory, but architectures combining both gauge equivariance and pseudofermion degrees of freedom have not previously been presented. 

This paper develops and presents flow architectures to model fermionic lattice gauge theories using pseudofermions. These architectures are based on ``joint models", where the action defining the target probability distribution is split into a ``marginal'' part, which depends only on the gauge variables, and a ``conditional'' part, which depends on the pseudofermionic variables given fixed gauge fields. The function of the conditional part is to efficiently estimate the fermion determinant. The new developments of this work are architectures to model the conditional component, and the introduction of a \linebreak {\it parallel transporter convolutional network}, which is the central piece of the architecture. In addition, it is outlined how standard approaches such as even/odd or Hasenbusch preconditioning can be combined with pseudofermion modeling.

Numerical demonstrations of the joint flow model architectures are provided in two toy gauge theories. One is the Schwinger model, which contains pseudo-Goldstone bosons, is confining, and has distinct topological sectors. The other is a 2D SU(3) gauge theory with $N_f=2$ flavors of fermions. While this system is topologically trivial, it shares the gauge group of QCD and can have large correlation lengths.

This paper is organized as follows. First the lattice actions and the pseudofermion approach are summarized in~\Cref{sec:lattice}. Flow-based sampling is reviewed in \Cref{sec:flowsreview}, joint models are constructed in \Cref{sec:joint}, the pseudofermion architectures are presented in  \Cref{sec:PTCL}, preconditioning is discussed in \Cref{sec:precons}, and 
a method to use multiple samples of pseudofermions given fixed gauge fields is explored in \Cref{sec:npf}.
Numerical demonstrations are provided in \Cref{sec:schwinger} for the Schwinger model and in \Cref{sec:SU3} for 2D SU(3). \Cref{sec:conclusion} presents a summary and conclusions. \Cref{app:ESSder} is a derivation of the scaling of the model quality with the number of pseudofermions at fixed gauge fields, and further details of the numerical exploration are provided in \Cref{app:details}. Finally, \Cref{app:hist} compares distributions of observables using two examples of flow models and Hybrid Monte Carlo (HMC) for sampling.

\section{Lattice Gauge Theories with Pseudofermions}
\label{sec:lattice}

The action of a lattice gauge theory with $N_f$ fermion degrees of freedom can be factorized into gauge and fermionic components as
\begin{equation}\label{eq:fullact}
    S(U,\psi, \bar{\psi}) = S_g(U) + S_f(U,\psi,\bar{\psi}) \ ,
\end{equation}
where the gauge links $U_\mu(x)$ are elements of the gauge group $\mathcal G$ and the fermion fields are collectively denoted as $\psi(x)$.
The Wilson discretization of the gauge part of the action is~\cite{Wilson:1974sk,Wilson:1975id}
\begin{equation}
    S_g(U) = -\frac{\beta}{N_c} \sum_x \sum_{\mu<\nu} \text{Re } \Tr P_{\mu\nu} \ ,
\end{equation}
where $\beta/2 N_c=1/g^2$ is the inverse of the squared gauge coupling, and $N_c$ refers to the dimension of the fundamental representation of the gauge group. The plaquette is defined as
\begin{equation}\label{eq:p_mu_nu}
P_{\mu \nu}(x):=U_{\mu}(x) U_{\nu}(x+\hat{\mu}) U_{\mu}^{\dagger}(x+\hat{\nu}) U_{\nu}^{\dagger}(x) \ ,
\end{equation}
where $\hat{\mu},\hat{\nu}$ denote unit vectors in direction $\mu,\nu$, respectively, and periodic boundary conditions are assumed. For 2D theories, the only contribution is $\mu=0$, $\nu=1$. 
The fermionic part of the action of a 2D gauge theory with $N_f$ degenerate flavors of fermions in the fundamental representation of $\mathcal G$ can be written as
\begin{equation}
    S_f(U,\psi,\bar{\psi}) = \sum_{f=1}^{N_f} \sum_{x,y} \bar{\psi}_f^{\gamma}(y) D[U](y,x)^{\gamma \alpha} \psi_f^{\alpha}(x) \ ,
\end{equation}
where $\psi_f^{\alpha}(x)$ denotes a fermion field with flavor $f$ and spin index $\alpha \in \{1,2\}$, and the gauge indices are kept implicit. 
The Wilson discretization~\cite{Wilson:1974sk,Wilson:1975id} of the lattice Dirac operator $D[U]$ is given by
\begin{align}\label{eq:dirac-wilson}
\begin{split}
    D[U](y, x)^{\gamma \alpha} &= \delta(y-x) \delta^{\gamma \alpha}  \\ 
    - ~ \kappa \sum_{\mu=0,1}& \, \Big\{  [1-\sigma_{\mu}]^{\gamma \alpha} U_{\mu}(y) \delta(y - x  + \hat{\mu})  \\ &
    +  [1+\sigma_{\mu}]^{\gamma \alpha} U^\dagger_{\mu}(y-\hat{\mu}) \delta(y - x - \hat{\mu}) \Big\} \ ,
    \end{split}
\end{align}
where $\sigma_{\mu}=(\sigma_x,\sigma_y)$, with $\sigma_{x,y}$ denoting the usual Pauli matrices, and $\kappa = 1/(4+2m_0)$, where $m_0$ is the bare fermion mass. Antiperiodic boundary conditions in the time direction are incorporated in the definition of the $\delta$ functions.

The full action, \Cref{eq:fullact}, is invariant under gauge transformations of the form
\begin{equation}
    U_\mu(x) \to \Omega(x) U_\mu(x) \Omega^\dagger(x+\hat{\mu}), \quad \psi(x) \to  \Omega(x) \psi(x) \ ,
\end{equation}
for any choice of $\Omega(x) \in \mathcal{G}$.  Observables are computed as expectations over field configurations,
\begin{equation}
    \langle \mathcal O \rangle = \frac{1}{\mathcal{Z}} \int DU D\psi D\bar \psi\, \mathcal O\, e^{-S(U,\psi, \bar{\psi})} \ ,
\end{equation}
where $p = e^{-S(U,\psi, \bar{\psi})}/ \mathcal Z$ plays the role of a probability density, with $\mathcal Z$ a normalization constant.

It is common practice to integrate out the Grassmann-valued fermionic degrees of freedom,
\begin{equation}\label{eq:intout}
   \frac{1}{\mathcal Z_f}  \int D\psi D\bar \psi\, e^{-S_f(U,\psi,\bar{\psi})} = (\det D[U])^{N_f} \ ,
\end{equation}
where $\mathcal Z_f$ is a normalization constant. This results in an effective action, which in the case of $N_f=2$ reads
\begin{equation}\label{eq:exactdet}
    S_{\rm eff}[U] = S_g(U) - \log \det D D^\dagger[U] \ ,
\end{equation}
where $\det D^2 = \det D D^\dagger$ due to $\gamma_5$-Hermiticity of $D$. This form of the action is given only in terms of gauge variables and can in principle be used for lattice calculations. However, the evaluation cost of the determinant scales poorly with the lattice volume. 
The scalable approach for state-of-the-art lattice QCD uses stochastic determinant estimators, in particular ones based on the following relation for positive definite matrices $M$:
\begin{equation}
    \det M = \frac{1}{(2\pi)^{N}}  \int d\phi \, e^{- \phi^\dagger M^{-1} \phi} \ ,
\end{equation}
where $\phi(x)$ are complex bosonic variables (the pseudofermions), $N$ is the number of pseudofermion variables, and $\int d\phi$ denotes integration over these variables. Adding these auxiliary variables, the $N_f=2$ theory can be represented with the following action:
\begin{align}
     S(U,\phi, \phi^\dagger) = S_g(U) + S_{\rm pf}(U,\phi, \phi^\dagger), \\  \text{ with }\, S_{\rm pf}(U,\phi, \phi^\dagger) = \phi^\dagger (D[U]D^\dagger[U])^{-1} \phi \ .
     \label{eq:pfaction}
\end{align}
In this form only the inverse of the Dirac operator applied to the pseudofermions is needed to evaluate the action, which can be computed with effectively linear cost scaling\footnote{For instance, the scaling of the conjugate gradient algorithm is $O(n_d \sqrt{k_n})$, where $n_d$ is the dimension of the matrix and $k_n$ is its condition number~\cite{10.5555/865018}.} with respect to the lattice volume. This comes at the cost of additional stochastic noise.
Several common variations on this approach, referred to as schemes for preconditioning the action, are described below.

\section{Flow models for pseudofermions}
\label{sec:flowsPF}

\subsection{Flow-based sampling}
\label{sec:flowsreview}

Normalizing flows~\cite{rezende2016variational,dinh2017density,JMLR:v22:19-1028} have proven to be a promising tool to mitigate critical slowing down and topological freezing in some lattice field theories~\cite{Albergo:2019eim,Kanwar:2020xzo,Albergo:2022qfi}. For an in-depth introduction to normalizing flows for lattice field theory, we refer the reader to Ref.~\cite{Albergo:2021vyo}. Here, we review the key concepts relevant for this work.

A ``flow'' is a diffeomorphism $f$ that is applied on a set of samples drawn from an easy-to-sample base (or prior) distribution, $r(z)$. The resulting configurations, $\varphi=f(z)$, are distributed according to the model distribution with density
\begin{equation}
    q(\varphi) = r(z)  \left|\det \frac{\partial f(z)}{\partial z}\right|^{-1} \ .
\end{equation}
The flow model is constructed with trainable parameters, which can be optimized to approximate a target probability distribution $p$, i.e., $q(\varphi)\simeq p(\varphi)$.

In the context of lattice field theory, configurations $\varphi$ are discretized quantum fields, and the target distribution is given by the Euclidean action of the theory, i.e.,~$p(\varphi) = e^{-S(\varphi)}/\mathcal Z$. Expressive transformations may be obtained by defining the flow $f$ as the composition of $n$ invertible layers
\begin{equation}
    f = g_1 \circ g_2 ... \circ g_n \ .
\end{equation}
Various architectures for the invertible layers $g_i$ tailored for different theories have been proposed in previous work, e.g., for scalar theories~\cite{Albergo:2019eim,Hackett:2021idh,Pawlowski:2022rdn}, Abelian and non-Abelian pure-gauge theories~\cite{Kanwar:2020xzo,Boyda:2020hsi}, and theories containing fermions~\cite{Albergo:2021bna}.

The optimization of a flow model proceeds by minimizing  a loss function that quantifies the difference between the model and target distributions. A common choice of loss is the Kullback-Leibler (KL) divergence~\cite{Kullback:1951}, see also \Cref{app:training}. A self-training optimization scheme can be used, in which the KL divergence is computed stochastically by drawing samples, or ``training data", from the model distribution. 
A useful measure of model quality during training is the Effective Sample Size per configuration (ESS), as defined, e.g., in Ref.~\cite{Albergo:2021vyo}:
\begin{equation}\label{eq:ESSgeneral}
    \text{ESS } = \frac{1}{N} \frac{\left( \sum_{i=1}^N p(\varphi_i)/q(\varphi_i)  \right)^2}{\sum_{i=1}^N \big (p(\varphi_i)/q(\varphi_i)\big)^2} \ ,
\end{equation}
where $N$ is the number of samples used to compute the estimate. Larger ESS implies better model quality, and $\text{ESS} \in [1/N\, ,\, 1]$.

Given a trained flow model, the most straightforward approach to sample from $p$ is to build a Markov chain using the independence Metropolis algorithm. Starting from some initial sample $\varphi$, the probability to accept an independent proposal $\varphi'$ generated by the flow model is defined as
\begin{equation}\label{eq:acc-rate}
\begin{aligned}
    A(\varphi \rightarrow \varphi') &= \min\left(1, \frac{p(\varphi')}{p(\varphi)}\frac{q(\varphi)}{q(\varphi')}\right) \ .
\end{aligned}
\end{equation}
Since each proposal from the flow model is independent from the previous one, autocorrelations in the Markov chain only arise from rejections in the Metropolis step. Higher quality models therefore result in less autocorrelation. Alternatively, one may use a reweighting procedure, where the weight of each configuration is given by
\begin{equation}
\label{eq:reweight-defn}
    w(\varphi) = p(\varphi)/q(\varphi)\ .
\end{equation}

\subsection{Joint models}
\label{sec:joint}

We build on Ref.~\cite{Albergo:2021bna} to define models for joint sampling of bosonic and pseudofermionic degrees of freedom.
Specifically, we construct ``joint autoregressive models’’ as described in Ref.~\cite{Albergo:2021bna}, which we simply refer to as ``joint models’’ in the following.
The fundamental idea is to factorize the probability distribution of \Cref{eq:pfaction} as
\begin{equation}
    p(U,\phi) = p(U) p(\phi|U) \ ,
\end{equation}
where each piece is defined as
\begin{equation} \label{eq:p-decomposition}
\begin{aligned}
    p(U) &\propto \det DD^\dagger[U]\,e^{-S_g(U)} \ , \\
    p(\phi|U) &\propto \frac{e^{-S_{\rm pf}(U,\phi,\phi^\dagger)}}{\det DD^\dagger[U]} \ ,
\end{aligned}
\end{equation}
referred to as the ``marginal'' and ``conditional'' distributions, respectively. As noted in Ref.~\cite{Albergo:2021bna}, sampling from $p(\phi | U)$ is straightforward, but computing the normalizing constant $\det D D^\dag$---needed for the ratios in  \Cref{eq:acc-rate,eq:reweight-defn}---is not.

We use two independent flow models to model the respective distributions: 
\begin{equation}
    q(U,\phi) = q(U) q(\phi|U) \simeq p(U) p(\phi|U) \ ,
\end{equation}
where the first component approximates the marginal distribution:
\begin{equation}
    q(U) = r_m(z) \left|\det \frac{\partial f_m(z)}{ \partial z}\right|^{-1} ,\ U = f_m(z) \ ,
\end{equation}
and the other models the conditional distribution:
\begin{equation}
     q(\phi|U) = r_c(\chi) \left|\det \frac{\partial f_c(\chi|U)}{\partial \chi}\right|^{-1}, \ \phi = f_c(\chi | U) \ ,
\end{equation}
where $r_m$ is taken to be uniform over the Haar measure and $r_c$ is a spherical Gaussian over all components of $\phi$. Note that the transformations defining the conditional flow, $f_c$, act only on the pseudofermions, that is, the gauge fields remain unchanged under the action of $f_c$. 

\begin{figure}[t!]
    \centering
    \vspace{-1cm}
    \includegraphics[width=\linewidth]{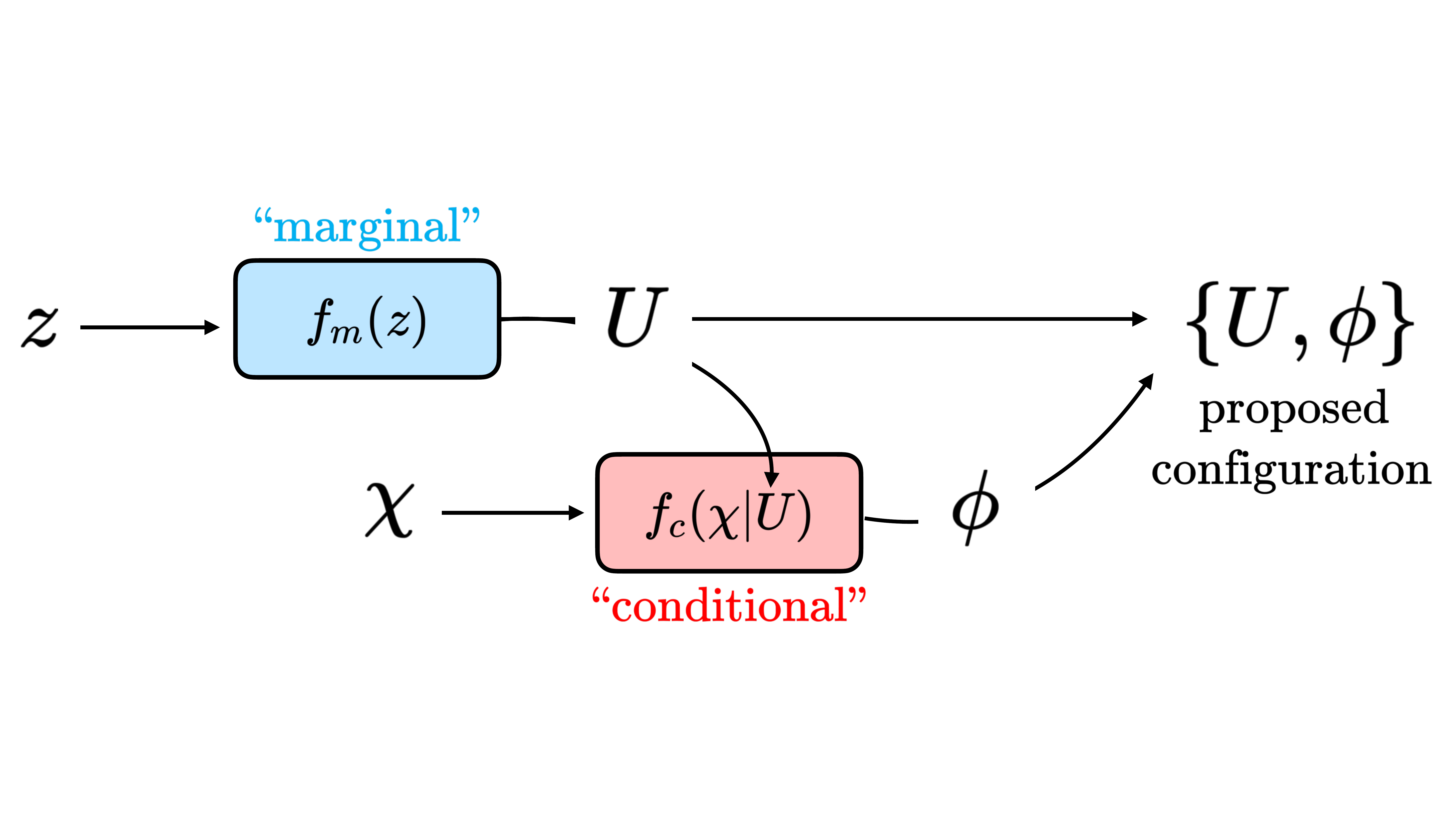}
    \vspace{-1.5cm}
    \caption{ Sketch of the workflow of a joint model. $z$ and $\chi$ are samples from the base distribution which are transformed to produce the gauge and pseudofermion fields, $U$ and $\phi$. $f_m(z)$ labels the flow model for the marginal distribution, and $f_c(z|U)$ that for the conditional one. The proposed configuration is distributed according to ${q(\phi, U) = q(U) q(\phi |U)}$.}
    \label{fig:diag}
\end{figure}

\Cref{fig:diag} sketches the sample generation workflow of a joint model. For $f_m(U)$ we use the gauge-equivariant layers described in Ref.~\cite{Boyda:2020hsi}, while $f_c(\phi|U)$ requires new technology which will be described in the following section; while pseudofermion architectures have already been presented in Ref.~\cite{Albergo:2021bna} for Yukawa theory, these did not treat pseudofermions coupled to gauge variables.  

To evaluate the quality of a joint model, one can use the ``joint ESS", defined by \Cref{eq:ESSgeneral} with ${q(\varphi) \to q(U,\phi)=q(U) q(\phi|U)}$ and ${p(\varphi) \to p(U,\phi)=p(U) p(\phi|U)}$. Another useful metric is the ``marginal ESS", which evaluates the quality of only the marginal model, cf.\ the target defined by \Cref{eq:exactdet}. This is obtained from  \Cref{eq:ESSgeneral} with $q(\varphi) \to q(U)$ and $p(\varphi) \to p(U)$.\footnote{Note that the marginal ESS requires the evaluation of the fermion determinant, and thus cannot be evaluated on larger volumes.}

\subsection{Gauge-equivariant architectures}
\label{sec:PTCL}

In this section, we describe how expressive gauge-equivariant transformations can be constructed for the conditional distribution of pseudofermions. In the approach of Ref.~\cite{Albergo:2021bna}, which treats pseudofermions that are not coupled to gauge degrees of freedom, the conditional flow $f_c$ is a function that maps an uncorrelated Gaussian distribution to a correlated one:
\begin{equation}
\label{eq:conditional-prior}
  r(\chi) \propto  e^{-\chi^\dagger \chi} \xrightarrow[]{\, f_c(\chi|U) \,} q(\phi|U) \propto e^{-\phi^\dagger (\tilde{D} \tilde{D}^\dagger[U])^{-1} \phi} \ .
\end{equation}
That is, the flow is a change of basis constructed as a linear transformation $\tilde{D}$ of the pseudofermions. Naturally, the choice $\tilde{D} = D$ would set $q(\phi | U) = p(\phi | U)$; however, evaluating the Jacobian of such a transformation would require evaluating the fermion determinant exactly, which is expensive. Instead the map $f_c$ can be constructed as a composition of simpler linear transformations, each with a tractable Jacobian.

In constructing such linear transformations, care must be taken to ensure the resulting map $f_c$ is gauge equivariant. A naive linear combination of pseudofermion variables at different lattice sites is not gauge equivariant, since pseudofermions at different positions transform differently. In order to construct a gauge-equivariant linear transformation, the pseudofermions must be parallel transported to a common point. For instance, the linear combination 
\begin{equation}
    \phi'(x) = a \phi(x) + b U_{\mu}(x) \phi(x+\hat \mu) \ ,
\end{equation}
with $a,b \in \mathbb{C}$, transforms as $\phi'(x) \to \Omega(x) \phi'(x)$, and thus defines an equivariant transformation.

We generalize this idea to introduce a {\it parallel transport layer}, $\mathbb{L}[U, \phi]$, which collects together the field at a given point $y$ with all of its nearest neighbors parallel transported to $y$, i.e.,
 \begin{equation}
     \mathbb{L}[U, \phi](y) = \{\phi(y),\phi^{x_1}(y), \ldots,\phi^{x_{2\times d}}(y)\} \ ,
 \end{equation}
where $x_i$ labels the coordinates of the $2\times d$ neighbors of $y$ (assuming $d$ space-time dimensions), and $\phi^{x_i}(y)$ is the pseudofermion $\phi(x_i)$ parallel transported to the location $y$. Note that antiperiodic boundary conditions must be applied when parallel transporting across the temporal boundary.
 
We use the parallel transport layer to form linear combinations of the pseudofermions in an equivariant way, similar to the technique developed in Ref.~\cite{favoni2020lattice}. Each input $\phi(y)$ has $K$ features, which in the this context indicate internal indices that are not gauge indices, e.g.~spinors in 2D have $K=2$ features. 
Thus, the concatenation of all vectors gathered by $\mathbb{L}[U, \phi](y)$ across their features outputs a new vector with $\tilde{K} = (2\times d+1)K$ features. This way, we define the {\it parallel transport convolution}, PTC$[U, \phi](y)$, as
 \begin{equation}\label{eq:PTCL}
     \text{PTC}[U, \phi]_{\alpha}(y) = \sum_{\gamma}M_{\alpha \gamma} \mathbb{L}[U, \phi]_{\gamma}(y) \ ,
 \end{equation}
where $\alpha,\gamma$ denote the feature indices, and the complex-valued matrix $M$ has dimensions $H \times \tilde{K}$. Here, $H$ is the number of features of the output of this layer.
In order to build expressive transformations and incorporate further information from the gauge field, one can parametrize  $M_{\alpha\gamma}$ as a neural network applied to the gauge field, for instance, a standard convolutional neural network with gauge invariant inputs. Finally, a {\it parallel transport convolutional network}, $\text{PTCN}[U, \phi](y)$, is defined as the composition of multiple $\text{PTCs}$  with $H=2$ (i.e., the number of spin components in 2D) in the last $\text{PTC}$ layer. The choice of $H$ at intermediate layers is arbitrary, and can be varied to increase expressivity. In a PTCN, the number of PTCs is denoted as  $n_\text{PT}$. We emphasize that PTCNs are linear operators on the input fields $\phi(x)$.

We are now in the position to define {\it pseudofermion layers}. To ensure that the Jacobian of the transformation can be computed efficiently, we perform variable partitioning into active, $\phi_a$, and frozen, $\phi_f$, degrees of freedom, encoded via a projector $\mathds{P}$ defined such that
\begin{equation}
    \phi_a(x) = \mathds{P}(x) \phi(x), \quad \phi_f(x) = (\mathds{1} - \mathds{P}(x)) \phi(x) \ .
\end{equation}
Then, a pseudofermion layer is defined by updating the active variables conditioned on the frozen ones as
\begin{equation}\label{eq:partitioning}
\begin{aligned}
    \phi'_a(x) &= A(x) \phi_a(x) +  \text{PTCN}[U, \phi_f](x), \\
    \phi'_f(x) &= \phi_f(x),
\end{aligned}
\end{equation}
where $A(x)$ is a site-dependent complex matrix in spin space (or is a scalar) that, like $M_{\alpha \gamma}$,  is also a function of gauge invariant combinations of links parametrized by a neural network.

The projector $\mathds{P}$ can be defined in different ways. One choice is ``spatial partitioning'', where the lattice sites are separated into active and frozen partitions, $x_a$  and $x_f$ respectively. This can be achieved, e.g., by using a checkerboard pattern. In this scheme, the projector takes the form
\begin{equation}\label{eq:spatial}
        \mathds{P}(x_a) = \mathds{1}_s, \quad \mathds{P}(x_f) = 0 \ ,
\end{equation}
and $\mathds{1}_s$ is the identity in the spinor indices. The Jacobian thus factorizes as
\begin{equation}
    \log J = 2 \sum_{x \in x_a}  \log |\det A(x)| \ ,
\end{equation}
where the factor of $2$ accounts for the complex numbers. 
Another choice of projector implements ``spin partitioning'', such that the projector is identical on all lattice sites and isolates a certain spinor component. In 2D, this can be defined as
\begin{equation}\label{eq:spin}
    \mathds{P}^\pm(x) =  \frac{1}{2}\left( \mathds{1}_s \pm \sigma_z\right) \ ,
\end{equation}
where $\sigma_z$ is the usual Pauli matrix, and $\mathds{P}^+$($\mathds{P}^-$) projects to the upper (lower) spinor component. In this case, the Jacobian of the transformation is also trivial to evaluate:
\begin{equation}
    \log J = 2 \sum_{x}  \log |A(x)| \ ,
\end{equation}
where $A(x)$ is now interpreted as a complex scalar that acts only on the active spin component.
 
A flow transformation can finally be defined as a composition of several pseudofermion layers. The parameters of the transformations of all layers---$M_{\alpha\gamma}$ in \Cref{eq:PTCL} and $A(x)$ in \Cref{eq:partitioning}---can be chosen to be the outputs of a ``context function'' built from neural networks.\footnote{This nested evaluation of neural networks to produce parameters for higher-level networks is similar to previous explorations of nested architectures in the machine learning community~\cite{sabour2017dynamic}.} This function only depends on the gauge links, and it can only take gauge-invariant inputs. We describe our choice in \Cref{app:details}. In the numerical demonstrations of \Cref{sec:results} we choose to alternate the choice of active and frozen variables such that all variables are updated with the same frequency. In addition, one can build a conditional flow using only spatial partitioning, only spin partitioning, or a combination of both.

\subsection{Joint models for preconditioned actions}
\label{sec:precons}

The Wilson-Dirac operator can have a large condition number due to the presence of small eigenvalues. This effect is enhanced when approaching the chiral limit of lattice gauge theories. Fortunately, this situation can be ameliorated by employing preconditioned actions, as is standard in state-of-the-art lattice QCD calculations using the HMC algorithm. Two common approaches are even/odd~\cite{DeGrand:1990dk} preconditioning and the Hasenbusch factorization~\cite{Hasenbusch:2001ne}. As detailed below, these schemes define modified target distributions cf. \Cref{eq:p-decomposition}, which require different architectures to model.

In the context of flow models, higher condition numbers imply more difficult target distributions, as the variance of the stochastic determinant estimator increases. This intuition is borne out in practice, as it can indeed be observed that it is more difficult to train flow models to model operators with larger condition number. Here, we describe how pseudofermion modeling can be combined with preconditioning techniques to mitigate this effect. As demonstrated in \Cref{sec:results}, this approach can be numerically advantageous.

\subsubsection{Pseudofermion models for even/odd preconditioning}
\label{sec:eo}

Even/odd (EO) preconditioning~\cite{DeGrand:1990dk} is a simple idea that reduces both the condition number and the number of degrees of freedom of the Dirac operator at almost no cost. It is based on the rearrangement of the Wilson-Dirac operator into the form:
\begin{equation}
  D = \begin{pmatrix}
  \mathbbm{1} & D_{eo}  \\
  D_{oe} & \mathbbm{1} \\
  \end{pmatrix} \ ,
\end{equation}
where $D_{oe}$ and $D_{eo}$ are the terms connecting nearest neighbors (``even" and "odd" sites). This way, the determinant can be calculated as
\begin{equation}
 \det D = \det \left(\mathbbm{1} - D_{eo} D_{oe} \right) \equiv \det D_\text{sc} \ ,
\end{equation}
where the subscript ``$\text{sc}$'' stands for Schur complement. Note that while $D \sim \mathbbm{1} + O(\kappa)$, the EO preconditioned Dirac operator is  $D_\text{sc} \sim \mathbbm{1} + O(\kappa^2)$, explaining the smaller condition number. The operator $D_\text{sc}$ acts on only even-site variables.

Pseudofermions can be used to estimate the EO preconditioned fermion determinant in the usual manner:
\begin{equation}\label{eq:eoPF}
     \det D D^\dagger \propto \int d\phi_e \, e^{-\phi_e^\dagger (D_\text{sc}^{\phantom{\dagger}} D_\text{sc}^\dagger)^{-1} \phi_e} \ ,
\end{equation}
where $\phi_e$ represents the pseudofermionic degrees of freedom defined only on the even sites of the lattice. 

In order to model the determinant in \Cref{eq:eoPF}, we must adapt our architecture to treat a pseudofermion field defined only on even sites. In practice, this can be done using architectures almost identical to those described in the previous section, by retaining odd-site variables but never updating their values from zero. Each PTCN may then populate these fields with values in intermediate states, so that they serve as additional ``working memory''. Then, the only architectural change must account for the fact that a single application of $\mathbb{L}[U, \phi]$ as defined in \Cref{eq:PTCL} connects only even sites to odd ones, and vice versa. To ensure that information is transferred across the $\phi_e(x)$ variables, one must apply always at least two $\text{PTC}[U, \phi]$ within the PTCN, i.e., $n_\text{PT}>1$. An alternative possibility is to modify the parallel transport layer, $\mathbb{L}[U, \phi]$, to parallel transport directly between even sites. Finally, note that spatial partitioning by checkerboarding is not well suited for EO-preconditioned targets, for which only even sites are defined. Instead, one may for example update even sites in every other row (or equivalently, column), alternating which rows/columns are updated from layer to layer; this approach is used in the numerical investigation of \Cref{sec:results}.

\subsubsection{Hasenbusch factorization in pseudofermion models}
\label{sec:hasenbusch}

Hasenbusch factorization~\cite{Hasenbusch:2001ne} is another common approach, and is a useful trick to separate the modes of the fermionic determinant. This can be achieved by factoring the determinant as
\begin{equation}\label{eq:hasenbusch}
    \det M = \left[ \frac{\det M}{\det \left( M + \mu \right)} \right]  \det \left( M + \mu \right) \ ,
\end{equation}
where $\mu>0$. Each of the factors in this equation are referred to as ``monomials". Each monomial can be estimated independently with separate pseudofermion fields:
\begin{equation}
\begin{aligned}\label{eq:mons}
    \frac{\det M}{\det \left(M + \mu \right)} & \propto  \int d\phi_0 \, e^{-\phi_0^\dagger (\mathbbm{1} + \mu M^{-1} ) \phi_0}, \\
{\det \left(M + \mu \right)} & \propto   \int d\phi_1 \, e^{-\phi_1^\dagger (M+\mu)^{-1}  \phi_1} \ ,
\end{aligned}
\end{equation}
associated with separate conditional target densities $p(\phi_0|U)$ and $p(\phi_1|U)$, respectively.
When $\mu=0$, the first monomial is trivial and the original problem is recovered. It can easily be seen that increasing $\mu$ numerically simplifies the evaluation of the second term of \Cref{eq:mons}, while making the evaluation of the first more difficult.

This procedure can be iterated $N_h$ times:
\begin{equation}\label{eq:hasenbuschN}
    \det M = \det \left( M + \mu_{N_h} \right) \prod^{N_h-1}_{i=0} \frac{\det \left( M + \mu_i \right)}{\det \left( M  + \mu_{i+1} \right)} \ ,
\end{equation}
where $\mu_0=0$ and $\mu_i<\mu_{i+1}$. In this equation, one has $N_h+1$ independent monomials for $N_h$ ``Hasenbusch steps''. In practice, the values of $\mu_i$ must be tuned to achieve the optimal performance;  the best choice is typically such that all monomials have similar average condition numbers. 

To combine this technique with flow models, one can use the architecture described in \Cref{sec:PTCL}, without modification, for each of the determinants to be estimated. This means constructing joint models containing multiple different conditional models:
\begin{equation} \label{eq:hasenbuchModel}
    q(U, \phi_0, \hdots, \phi_{N_h+1}) = q(U) \prod^{N_h+1}_i q(\phi_i | U) \,,
\end{equation}
where $q(\phi_i | U)$ is the density of the flow model for the monomial $i$. 
The ESS for the resulting joint model can be computed using \Cref{eq:ESSgeneral} with the weight factor
\begin{equation}
    w(U, \phi_0, \ldots, \phi_{N_h+1}) = \frac{p(U)}{q(U)} \prod_i^{N_h+1} \frac{ p(\phi_i|U)}{q(\phi_i|U)} \ .
\end{equation}

Furthermore, the EO preconditioning introduced in \Cref{sec:eo} can be easily combined with the Hasenbusch factorization, as is common in state-of-art lattice QCD studies, by simply replacing $M = D D^\dagger \to D_\text{sc}  D_\text{sc}^\dagger$ in \Cref{eq:hasenbuschN}. Combinations of both techniques will be the default approach used in the numerical demonstrations discussed in \Cref{sec:results}.

\subsection{Improving the stochastic determinant estimate}
\label{sec:npf}

In order to use joint models more efficiently, one can draw multiple pseudofermion samples for fixed gauge fields. This provides more precise estimators of the determinant of the Dirac operator.
The resulting improved weights can be used with the pseudo-marginal Markov-chain Monte Carlo algorithm~\cite{Andrieu:2009,Albergo:2021bna} to provide better statistical performance.
The procedure is outlined below.

For a fixed gauge field $U$, the determinant of $M(U)$ can be estimated as
\begin{equation}
\begin{aligned}
    \det  M(U)&=\int d \phi\, e^{-S_{\rm pf}(U,\phi,\phi^\dagger)}\\&=\int d \phi\, q(\phi | U) \frac{e^{-S_{\rm pf}(U,\phi,\phi^\dagger)}}{q(\phi | U)} \\
    &=\left\langle\frac{e^{-S_{\rm pf}(U,\phi,\phi^\dagger)}}{q(\phi | U)}\right\rangle_{q} \ ,
\end{aligned}
\end{equation}
where $\langle \cdots \rangle_q$ denotes the average over samples distributed as $q(\phi|U)$. This way, one can define the weight factor for the conditional part using $N_{\rm{pf}}$ pseudofermion samples as
\begin{equation}
    {w}_{N_{\rm{pf}}}(\{\phi\} | U) =   \frac{1}{N_{\rm{pf}}} \sum_{i=1}^{N_{\rm{pf}}} \frac{{p}(\phi^{(i)} | U)}{q(\phi^{(i)} | U)} \ . \label{eq:condw}
\end{equation}
Combining this with the weight factor of the marginal model, the weight factor of a single gauge configuration $U$ and corresponding set of $N_{\rm{pf}}$ pseudofermion samples $\{\phi\}$ can be defined as 
\begin{equation} 
    \begin{aligned}
    {w}_{N_{\rm{pf}}}( U ) &= \frac{p(U)}{q(U)} \times {w}_{N_{\rm{pf}}}(\{\phi\} | U) \\
    &= \frac{1}{N_{\rm{pf}}} \sum_{i=1}^{N_{\rm{pf}}} \frac{{p}(\phi^{(i)}, U)}{q(\phi^{(i)}, U)} \ .
    \end{aligned}
\end{equation}
This way, one can use the weight factor ${w}_{N_{\rm{pf}}}( U )$ averaged over several pseudofermion draws for a Metropolis accept-reject step, which leads to higher acceptance rates. In the limit of infinitely many pseudofermion samples, this converges to using the marginal flow model with an exact evaluation of the determinant of the Dirac operator. 
As a metric, one can also define the ESS for several pseudofermions:
\begin{equation}\label{eq:ESSnpf}
    \text{ESS}({N_{\rm{pf}}}) = \frac{1}{N} \frac{\left(\sum_{i=1}^N {w}_{N_{\rm{pf}}}( U^{(i)} ) \right)^2 }{\sum_{i=1}^N {w}_{N_{\rm{pf}}}( U^{(i)} )^2} \ .
\end{equation}
As demonstrated in \Cref{app:ESSder}, this depends on the number of pseudofermion draws as
\begin{equation}
    \text{ESS}(N_{\rm{pf}}) = \frac{\text{ESS}(\infty)}{1+\frac{C}{N_{\rm{pf}}}} \ , \label{eq:ESSscalingNPF}
\end{equation}
where $C$ is a constant, and $\text{ESS}(\infty)$ coincides with the marginal ESS.

Finally, when using this procedure with the Hasenbusch factorization, one must draw $N_{\rm{pf}}$ independent pseudofermion samples for each monomial, for a total of $N_{\rm{pf}} (N_h + 1)$. The weight factor for each monomial, ${w}_{N_{\rm{pf}}}(\{\phi_i\} | U)$, will be computed as in \Cref{eq:condw}, and the overall weight factor ${w}_{N_{\rm{pf}}}( U )$ will result from multiplying all the weight factors from each monomial with the marginal part:
\begin{equation}
    {w}_{N_{\rm{pf}}}( U ) = \frac{p(U)}{q(U)} \times \prod^{N_h+1}_i {w}_{N_{\rm{pf}}}(\{\phi_i\} | U) \ .
\end{equation}

\section{Numerical Demonstration}
\label{sec:results}

This section presents numerical examples of applications of the joint modeling approach to sampling gauge field configurations for two 2D toy theories. A subsection is dedicated to each of the Schwinger model and SU(3) in 2D. More details about the numerical experiments can be found in \Cref{app:details}, and examples of distributions of values for observables using two of these models for sampling in \Cref{app:hist}.

\begin{figure}[t!]
    \centering
    \includegraphics[width=\linewidth]{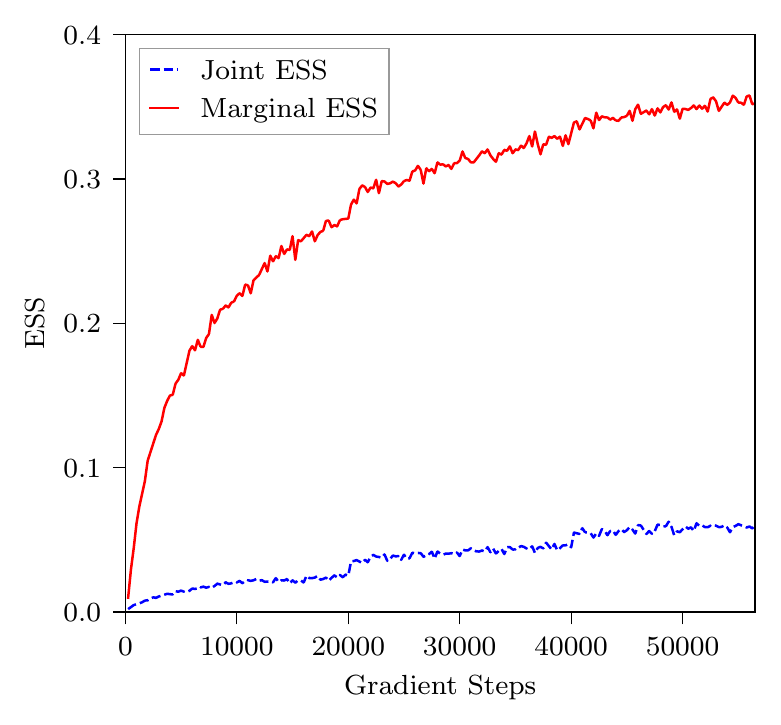}
    \caption{Joint and marginal ESS for a flow for the Schwinger model with $L=16$ and model parameters and flow model architecture as described in the text. The jumps in ESS are a result of learning-rate scheduling; see Appendix~\ref{app:details} for details of the training hyperparameter choices.}
    \label{fig:ESS16}
\end{figure}

\subsection{\texorpdfstring{$N_f=2$}{Nf=2} Schwinger model}
\label{sec:schwinger}

The Schwinger model, that is, 2D QED, shares some features with QCD~\cite{Schwinger:1962tp,Coleman:1975pw} and is often used as a testbed for new algorithms for lattice gauge theories~\cite{Smit:1987fq,Dilger:1992yn,Dilger:1994ma,Durr:2012te,Funcke:2019zna,Butt:2019uul,Banuls:2019bmf,Albandea:2021lvl,Finkenrath:2022ogg,Hartung:2021wqg,Eichhorn:2021ccz}. In the context of flow-based sampling and related approaches, this theory has already been investigated in Refs.~\cite{Albergo:2022qfi,Finkenrath:2022ogg}. However, these works used the exact determinant action of \Cref{eq:exactdet}, while here we use the pseudofermion approach.

For the numerical demonstrations, we use the following parameters for the action: $\beta=2.0$ and $\kappa=0.265$ (corresponding to a bare fermion mass of $m_0=-0.113208$). The pseudoscalar mass is $aM_{PS} \sim 0.35$. We implement an architecture using the spatial variable partitioning presented in \Cref{eq:spatial}. 
Furthermore, we use a regulator of the Dirac operator during training in the form of $(D D^\dagger + \mu_0)$, with $\mu_0 = 10^{-5}$. 
We use this regulated target only during training, for stability. During sampling and when computing observables, we use the unregulated operator and no additional approximation is induced by this procedure.

\subsubsection{Example of joint model}
\label{sec:exampleL16}

\begin{figure}[t!]
    \centering
    \includegraphics[width=\linewidth]{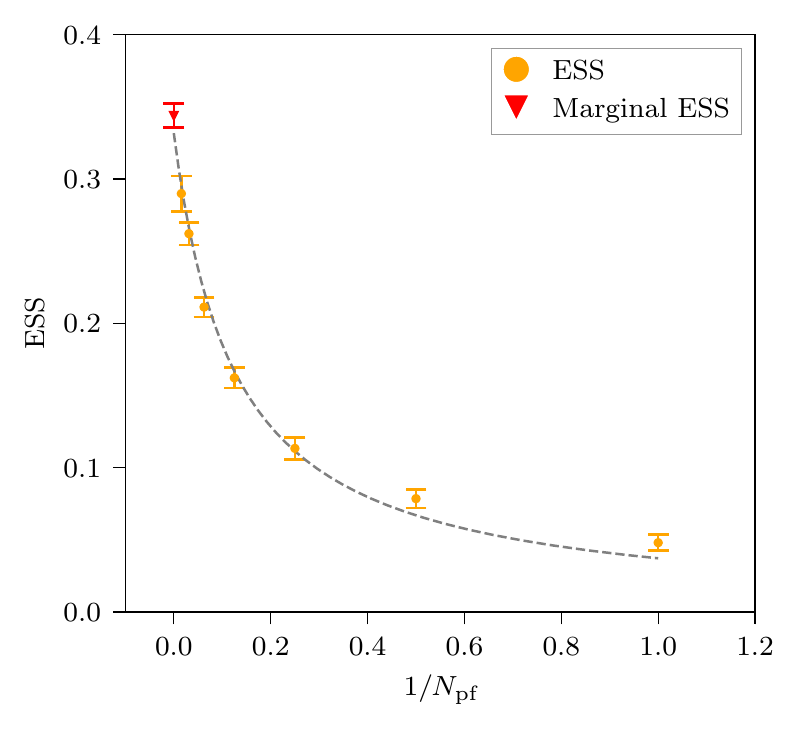}
    \caption{Scaling of the ESS as defined in \Cref{eq:ESSnpf} as a function of $1/{N_{\rm{pf}}}$, where $N_{\rm{pf}}$ is the number of pseudofermions drawn with fixed gauge fields. Each orange point corresponds to a factor of 2 increase in $N_{\rm{pf}}$, from 1 to 64. The red triangle corresponds to the marginal ESS of the model. Uncertainties are computed by using 10 independent estimations of the ESS on batches of 10240.
    \vspace{-1em}}
    \label{fig:morePFs}
\end{figure}

This section presents the results of a joint model trained for lattice extent $L=16$. For the pseudofermionic target, we use EO preconditioning with three Hasenbusch iterations and ${(\mu_1, \mu_2, \mu_3)=(0.001,0.01,0.05)}$. We train using a batch size of 1344 configurations per gradient step, and use models with 32 pseudofermion layers. In total, there is one flow model to generate the gauge links, and four independent models for the pseudofermions, one for each Hasenbusch monomial (some weights are shared in the context function of these models, as discussed in \Cref{app:details}). We train the joint model as a whole, optimizing all five components simultaneously. The training curve for the joint and marginal ESS is shown in \Cref{fig:ESS16}. As can be seen, a $\sim 5\%$ joint ESS is obtained, while the marginal ESS is $\sim 35 \%$. When employed for sampling using the independence Metropolis algorithm with a batch of 20k configurations, this flow model provides an acceptance rate of $\sim 18 \%$.

A way to improve the sampling quality is to follow the procedure outlined in \Cref{sec:npf}, that is, drawing more pseudofermion samples for fixed gauge fields. This is shown in \Cref{fig:morePFs}, where the ESS as a function of the number of pseudofermion draws, $N_{\rm{pf}}$, is shown. In that figure, from right to left, every point corresponds to an increase by a factor of 2 in $N_{\rm{pf}}$. It can be clearly seen that the ESS increases with larger $N_{\rm{pf}}$ and approaches the marginal ESS. To guide the eye, a fit to \Cref{eq:ESSscalingNPF} is also shown. 
This method is a promising option for increasing the statistical precision of measured observables when using flow-based sampling, as it requires only additional pseudofermion sampling without modifying or generating additional gauge fields. Indeed, training can be done with a small $N_{\rm{pf}}$, and then $N_{\rm{pf}}$ may be increased arbitrarily during evaluation. 
Because statistical quality can be improved without generating additional gauge field samples, the approach can be particularly advantageous in situations where the evaluation of observables dominates the computational cost. This is in fact the case in many lattice QCD applications.

\subsubsection{Effect of preconditioning}
\label{sec:effectprecons}

\begin{figure*}[t!]
\vspace{0.5cm}
     \centering
     \subfloat[Joint ESS\label{fig:jointESS}]{%
     \includegraphics[width=0.49\textwidth]{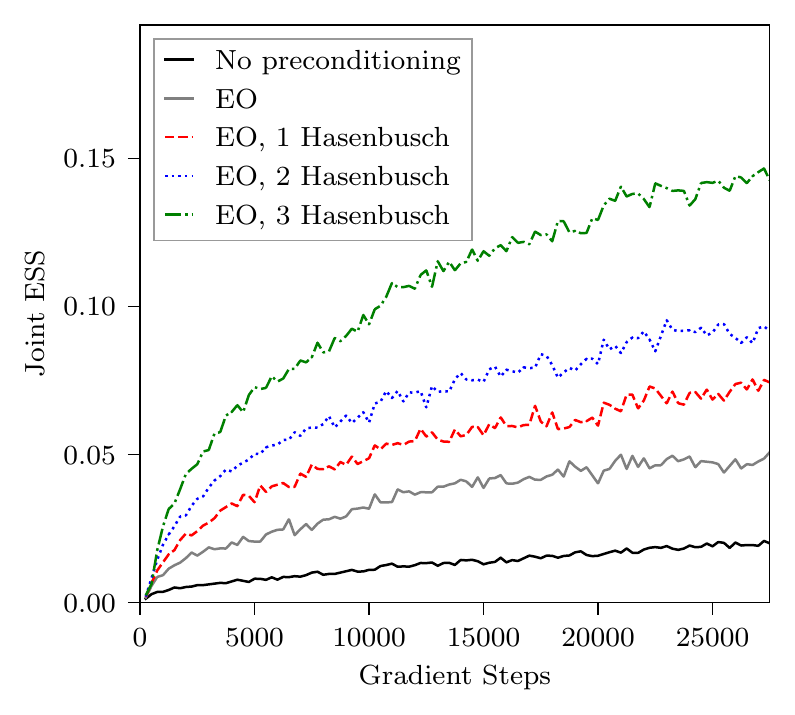}
    }
    \hfill
    \subfloat[Marginal ESS\label{fig:margESS}]{%
     \includegraphics[width=0.49\textwidth]{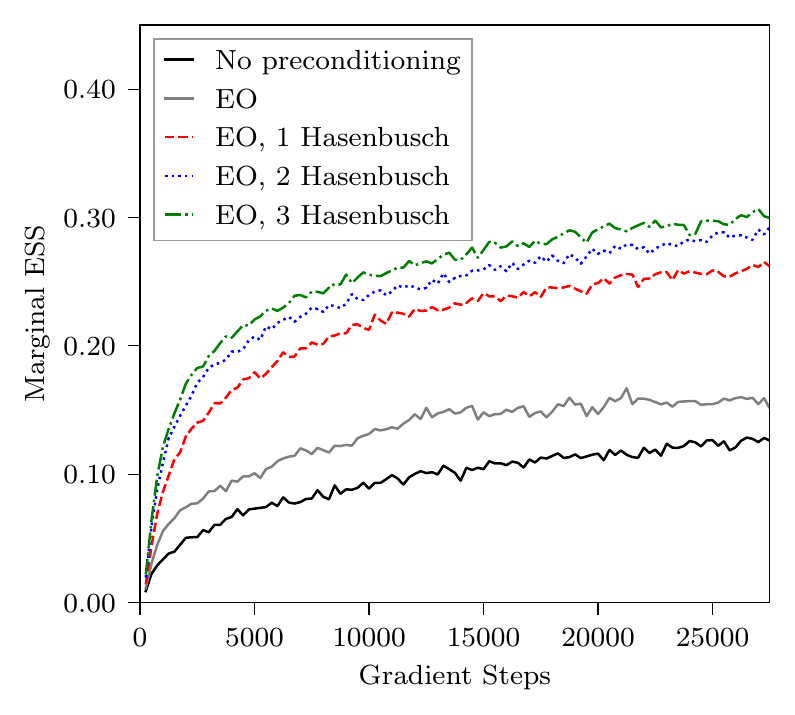}
    }

    \caption{Joint ESS (left) and marginal ESS (right) for flow models trained for the Schwinger model at $L=8$, $\beta=2$ and $\kappa=0.265$ with five different choices of preconditioning. 
    From bottom to top, the curves show results with no preconditioning, using EO preconditioning, and one, two, or three iterations of the Hasenbusch factorization. The training procedure and the marginal architectures (with 24 gauge layers) are the same for all curves. All conditional models have 24 pseudofermion layers. All Hasenbusch monomials are modeled by separate conditional models.
    The ESS is evaluated with batch size 4096.}
    \label{fig:ESS}
       \vspace{0.1cm}
\end{figure*}

This section provides a numerical demonstration of how different preconditioning schemes interact with flow model quality for the Schwinger model. For these tests, we use a similar model architecture as detailed in the previous section, with 24 pseudofermion layers and the same target parameters as in the previous example, but at a smaller lattice extent, $L=8$. \Cref{fig:ESS} shows the joint ESS (left) and marginal ESS (right) of models for a set of five different pseudofermion targets: (i) no preconditioning, (ii)  even/odd, (iii) even/odd and Hasenbusch with one monomial, $\mu_1=0.001$, (iv) even/odd and Hasenbusch with two monomials, $(\mu_1, \mu_2)=(0.001,0.01)$, and (v) even/odd and Hasenbusch with three monomials, ${(\mu_1, \mu_2, \mu_3)=(0.001,0.01,0.05)}$.

Clearly, EO preconditioning results in significantly increased ESS, as is to be expected from the reduced number of pseudofermion variables. This improvement is also visible in the marginal ESS, which suggests better propagation of information about fermionic effects to the marginal model. 
Another source of intuition about target complexity is condition number.
The unpreconditioned $D D^\dagger$ operator has an average condition number of $\sim 960$, while that with EO preconditioning is $\sim 195$ for test ensembles of 40k gauge fields generated from each flow model. However, EO preconditioning does not prevent the occasional appearance of very large condition numbers---in test ensembles with 40k configurations, condition numbers as large as $O(10^5)$ are found both with and without EO. 

Considering the combination of EO with Hasenbusch factorization, one might naively expect the ESS to decrease with more Hasenbusch steps since adding monomials increases the number of pseudofermion variables to model. However, both the joint and marginal ESS improve, with more improvement if more Hasenbusch steps are used. Therefore, the Hasenbusch-preconditioned target must be significantly easier to model, which is confirmed by considering the condition number of the matrices involved; for the case of three factorization steps, in an ensemble of 40k configurations generated from this model the average condition number is $O(10)$ in all four monomials, and the maximum value is less than $100$ in all cases. We conclude that while using Hasenbusch requires additional conditional flows and thus larger models overall---with the corresponding increased memory and computing cost---it may provide a systematic approach for increasing the sampling quality.

\subsection{Application to SU(3) in 2D with \texorpdfstring{$N_f=2$}{Nf=2}}
\label{sec:SU3}

This section details a demonstration of the gauge-equivariant joint models to sample a 2D SU(3) theory with two fermion flavors. This toy model also has some similar features to QCD, such as the gauge group, confinement, and light meson-like bound states. It is however intrinsically different than QCD, since 2D SU($N$) theories are topologically trivial. For this investigation we choose the following parameters: $\beta=6$, $\kappa=0.265$ (corresponding to $m_0 = -0.113208$), and lattice volume $L^2$ with $L=16$. Note that for this value of $\beta$, the bare gauge coupling is $g^2_0 = 6/\beta = 1$, and so it is in the physically relevant weak-coupling region. This choice of $\kappa$ results in a pseudoscalar mass of $aM_\pi \simeq 0.72$. 

We model a target preconditioned with EO and one step of the Hasenbusch factorization with $\mu_1=0.3$. We find that more steps are not necessary, since the Dirac operator has lower condition number than in the Schwinger model. We build the pseudofermion flows alternating spatial and spin variable partitioning, which provides better models than using either scheme alone. We use the PTCN as described in \Cref{sec:PTCL} with $n_\text{PT}=6$, and $H=4$ in the intermediate layers. Increasing the values of the latter parameters does not lead to substantial improvements, while increasing the cost. 

\begin{figure}[t!]
    \centering
    \includegraphics[width=\linewidth]{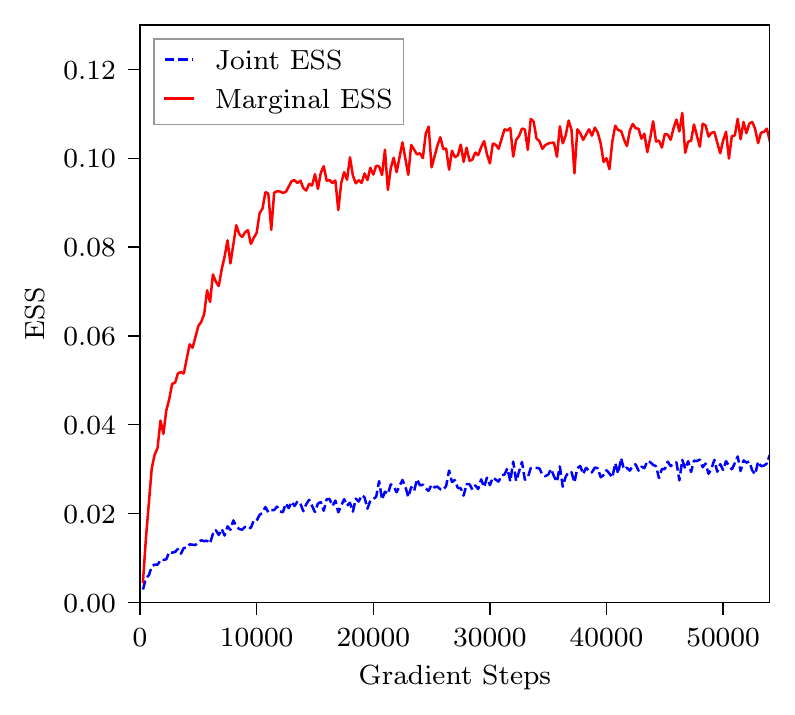}
    \caption{Marginal and joint ESS of a flow model trained as described in the text for a 2D SU(3) theory with $L=16$.}
    \label{fig:SU316}   
\end{figure}

\Cref{fig:SU316} shows the joint and marginal ESS along training. A joint ESS of $\sim 3 \%$ is achieved after 54k steps of training. When using this model in sampling, this yields an acceptance rate of $\sim 8\%$.

\section{Conclusion and Outlook}
\label{sec:conclusion}

This work presents a crucial development for scalable flow-based modeling of gauge theories with fermion content. Specifically, we have introduced gauge-equivariant architectures for the pseudofermion approach to the fermion determinant. These architectures are used in the context of joint models, where the probability distribution is split into the marginal and conditional part---the former depends on the gauge variables, and the latter on the pseudofermions with fixed gauge fields. The central piece of technology employed in this work to model the conditional part is the {\it parallel transport convolutional network}. A PTCN is the composition of several {\it parallel transport convolutions}, which compute linear combinations of pseudofermions that have been parallel transported to a common site. This is described in depth in \Cref{sec:PTCL}.

An important observation is that a large condition number in the Dirac operator makes models for the conditional target harder to optimize. Therefore, we argue that preconditioners should also be used in the context of flow models for pseudofermions. In particular, we describe how the new pseudofermion layers can be easily adapted to be used in conjunction with even/odd preconditioning and the Hasenbusch factorization.

Numerical implementations of the approach are provided in 2D gauge theories, namely the Schwinger model and 2D SU(3). These investigations demonstrate the advantages of preconditioning to obtain higher-quality models, and show how using more pseudofermion draws can systematically improve statistical performance.

The ideas of this work can be applied to four spacetime dimensions without formal complications, enabling flow-based sampling of theories of phenomenological interest such as QCD~\cite{Abbott:2022hkm}.
In practice, however, significant further engineering will be required to design and train architectures well-suited to treating the more complex structures that arise in higher dimensions and in theories with more complex topological features.
As discussed at length in Ref.~\cite{scaling}, determining whether the approach presented here will provide a viable approach to QCD sampling at state-of-the-art parameters is a critical question which will require direct exploration.

\section*{Acknowledgements}
RA, GK, DCH, FRL, and PES are supported in part by the U.S.\ Department of Energy, Office of Science, Office of Nuclear Physics, under grant Contract Number DE-SC0011090. DCH, FRL and PES thank the Institute for Nuclear Theory at the University of Washington for its kind hospitality and stimulating research environment. This research was supported in part by the INT's U.S. Department of Energy grant No. DE-FG02-00ER41132. PES is additionally supported by the National Science Foundation under EAGER grant 2035015, by the U.S.\ DOE Early Career Award DE-SC0021006, by a NEC research award, and by the Carl G and Shirley Sontheimer Research Fund. GK is additionally supported by the Schweizerischer Nationalfonds. KC and MSA are supported by the National Science Foundation under the award PHY-2141336. MSA thanks the Flatiron Institute for their hospitality. DB is supported by the Argonne Leadership Computing Facility, which is a U.S. Department of Energy Office of Science User Facility operated under contract DE-AC02-06CH11357. This work is funded by the Deutsche Forschungsgemeinschaft (DFG, German Research Foundation) under Germany's Excellence Strategy EXC 2181/1 - 390900948 (the Heidelberg STRUCTURES Excellence Cluster), the Collaborative Research Centre SFB 1225 (ISOQUANT), and the U.S.\ National Science Foundation under Cooperative Agreement PHY-2019786 (The NSF AI Institute for Artificial Intelligence and Fundamental Interactions, \url{http://iaifi.org/}). This work is associated with an ALCF Aurora Early Science Program project, and used resources of the Argonne Leadership Computing Facility, which is a DOE Office of Science User Facility supported under Contract DEAC02-06CH11357. The authors acknowledge the MIT SuperCloud and Lincoln Laboratory Supercomputing Center~\cite{reuther2018interactive} for providing HPC resources that have contributed to the research results reported within this paper. Numerical experiments and data analysis used PyTorch~\cite{NEURIPS2019_9015}, JAX~\cite{jax2018github}, Haiku~\cite{haiku2020github}, Horovod~\cite{sergeev2018horovod}, NumPy~\cite{harris2020array}, and SciPy~\cite{2020SciPy-NMeth}. Figures were produced using matplotlib~\cite{Hunter:2007}.

\appendix

\section{ESS with multiple pseudofermions samples}
\label{app:ESSder}

For any gauge field $U$ we can estimate the conditional reweighting factor $w_{N_{\rm{pf}}}(\{\phi\}|U)$ with $N_{\rm{pf}}$ samples from the pseudofermion distribution as in \Cref{eq:condw}. Asymptotically,
\begin{equation}\begin{aligned}
    \text{ESS}^{-1}(N_{\rm{pf}}) 
    &= \braket{w(U,\{\phi\})^2}_{q(U,\{\phi\})}
    \\
    & \equiv \int dU \, q(U) \, w(U)^2 \, \epsilon^{-1}_{N_{\rm{pf}}}(U) \ ,
\end{aligned}\end{equation}
where
\begin{equation}\begin{aligned}
    \epsilon^{-1}_{N_{\rm{pf}}}(U) 
    &= \int \left( \prod_{i=1}^{N_{\rm{pf}}} d\phi^{(i)} \,q(\phi^{(i)} |U) \right) w_{N_{\rm{pf}}}(\{\phi\}|U)^2
    \\ 
    &\equiv \left\langle w_{N_{\rm{pf}}}(\{\phi\}|U)^2 \right\rangle_{\phi^{(1)} \cdots \phi^{({N_{\rm{pf}})}}} 
\end{aligned}\end{equation}
and the second line is simply defining compact notation for the expectations over $q(\cdot|U)$.
Using $\braket{w}=1$,
\begin{equation}\begin{aligned}
    \epsilon^{-1}_{N_{\rm{pf}}}(U)
    &= \frac{1}{N_{\rm{pf}}^2} \left \langle { 
        \sum_i w_i^2 + \sum_{i \neq j} w_i w_j
} \right \rangle_{\phi^{(1)} \cdots \phi^{({N_{\rm{pf}})}}} 
    \\
    &= \frac{1}{N_{\rm{pf}}} \braket{w(\phi|U)^2}_{q(\phi|U)} + \frac{N_{\rm{pf}} - 1}{N_{\rm{pf}}}
    \\
    &\equiv 1 + \frac{X(U)}{N_{\rm{pf}}} \, ,
\end{aligned}\end{equation}
where $w_i \equiv w(\phi^{(i)}|U)$, and in the last line we have isolated the $U$ and $N_{\rm{pf}}$ dependence.
Inserting back into the expression for the full ESS and evaluating the integrals, we obtain
\begin{equation}\begin{aligned}
    \text{ESS}^{-1}(N_{\rm{pf}})
    &= \int dU q(U) w(U)^2 \left[ 1 + \frac{X(U)}{N_{\rm{pf}}} \right]
    \\
    &= \text{ESS}^{-1}(\infty) + \frac{C'}{N_{\rm{pf}}},
\end{aligned}\end{equation}
where $\text{ESS}(\infty)$ is the marginal ESS.
The result is identical to \Cref{eq:ESSscalingNPF} up to algebraic manipulations and redefinition of $C'$.

\section{Further details of numerical experiments}
\label{app:details}

This section provides additional details of the architecture and training scheme for the numerical implementations of the flow models described in \Cref{sec:results}. 

\subsection{Training and optimization}
\label{app:training}

The self-training scheme uses a loss function that is a stochastic estimate of the Kullback-Leibler divergence~\cite{Kullback:1951} computed with $q$-distributed samples generated by the model,
\begin{equation}\label{eq:loss}
\begin{aligned}
    D_{\mathrm{KL}}(q||p) 
    &= \int dU \, q(U) \log \frac{q(U)}{p(U)} \\
    &\approx \frac{1}{B} \sum_{i=1}^B \left[ \log q(U_i) + S(U_i) \right] + (\text{const}) \ ,
\end{aligned}
\end{equation}
where $B$ is the batch size (i.e.,~number of field samples generated for each gradient step). The constant does not affect optimization, and it is ignored.
We have two independent implementations of the experiments, one using PyTorch and the other using JAX~\cite{jax2018github}. In the PyTorch setup, we train using the AdamW optimizer~\cite{loshchilov2017decoupled}.  The rest of the parameters correspond to the Pytorch 1.12 default. We initialize the weights using ``Xavier normal'' Pytorch initialization with $\text{gain}=0.5$. In the JAX setup we use the Adam optimizer with gradient clipping and the default Haiku~\cite{haiku2020github} initialization with variance scaling to make the flows closer to the identity map at initialization. The results from both implementations are consistent, but the specific models shown in this paper are trained using the PyTorch implementation.

\subsection{Context function for the pseudofermion layers}

The linear transformation of pseudofermion variables in each PTC is parameterized by the outputs of a context function. To preserve linearity in $\phi(x)$, this context function cannot depend on the pseudofermions. Moreover, using only use gauge-invariant inputs ensures the gauge equivariance of the transformation. In the numerical implementation in this work, in order to respect translation symmetry, we build the context function from 2D convolutions with periodic boundary conditions. The first part of the context function is shared by all the pseudofermion layers, and maps $N_i$ input channels to $N_\text{hidden}$ hidden channels. If the Hasenbusch factorization is used, all monomials also share this first part. The second part of the convolution is specific for each pseudofermion layer, and maps the $N_\text{hidden}$ hidden channels to $N_o$ channels. In intermediate convolutions the ELU~\cite{2015arXiv151107289C} activation is applied, and $\tanh$ is used in the final one.

We use four input channels to the convolution:
\begin{equation}
    \text{Re} \Tr P_{01}(x), \text{Im} \Tr P_{01}(x), I_0(x), I_1(x) \ ,
\end{equation}
where $P_{01}$ is the plaquette, and
\begin{equation}
I_i(x) = x_i \text{ mod } 4
\end{equation}
is a constant input with periodicity mod 4. We find that including $I_i$ improves the quality, but is nonessential. The number of output channels will depend on the nature of the pseudofermion layer. First, for each PTC in the PTCN, $2\times 5 \times H \times H'$ real numbers are required to parametrize a generic complex $M_{\alpha\gamma}(x)$ in \Cref{eq:PTCL}. Moreover, $8$ complex numbers are required for $A(x)$ in \Cref{eq:partitioning} if spatial partitioning is used. The neural network outputs are taken as the real and imaginary parts of the entries of all complex matrices, with the exception of $A(x)$ in \Cref{eq:partitioning} to which we add the identity.

\subsection{Hyperparameters for the Schwinger model}

First, we list the common hyperparameters for all the flow models in \Cref{sec:schwinger}. For the U(1) gauge layers, we use the architecture and masking pattern described in the Appendix of Ref.~\cite{Albergo:2022qfi}. The differences with respect to that work are the number of gauge layers and the number of hidden channels in the context function (here we use 32). We use $\epsilon=0.02$ in the AdamW optimizer. The PTCNs have $n_\text{PT}=6$ when EO is used, and $n_\text{PT}=3$ otherwise. ${N_\text{hidden}=16}$ is used in the context function for the pseudofermion layers. $H=2$ is used in the intermediate PTCs. These models are constructed only with a spatial partitioning of the pseudofermion variables.

In addition, other hyperparameters vary between \Cref{sec:exampleL16} and \Cref{sec:effectprecons}.
\begin{enumerate}
    \item The results shown in \Cref{fig:ESS16} and \Cref{sec:exampleL16} are generated from a model trained with batch size $B=1344$, and $L=16$. The model has 32 pseudofermion layers, as well as 32 gauge layers. The initial learning rate (LR) is $\eta= 7\cdot 10^{-4}$, which is reduced by a factor of $0.5$ every 20k gradient steps. 
    \item All the models in \Cref{fig:ESS} and \Cref{sec:effectprecons} have $B=4096$, $L=8$, 24 gauge layers, and 24 pseudofermion layers for each monomial. The LR starts from $\eta= 5\cdot 10^{-4}$, and decays by a factor of $0.5$ every 10k gradient steps. 
\end{enumerate}

\subsection{Hyperparameters for SU(3) in 2D}
The flow model described in \Cref{sec:SU3} has 48 gauge layers and 48 pseudofermion layers in each of the two monomials. The PTCNs have $n_\text{PT}=6$, and $H=4$ is used in the intermediate PTCs. $N_\text{hidden}=16$ is used in the context function for the pseudofermion layers. Spatial and spin partitioning are alternated in the model. 
For the SU(3) gauge layers, we use the architecture and masking pattern described in Ref.~\cite{Boyda:2020hsi}. The context function for the gauge part has two intermediate standard convolutions, with 32 hidden channels.
For training, we use $\epsilon=0.01$ in the AdamW optimizer. We use a batch size of 576. The LR scheduling is as follows: beginning from $\eta= 5\cdot 10^{-4}$, the LR is decayed by a factor of $0.5$ every 10k gradient steps.

\section{Histograms with observables}
\label{app:hist}

\begin{figure*}[t!]
     \centering
     \subfloat[Average plaquette in U(1).]{%
     \includegraphics[width=0.48\textwidth]{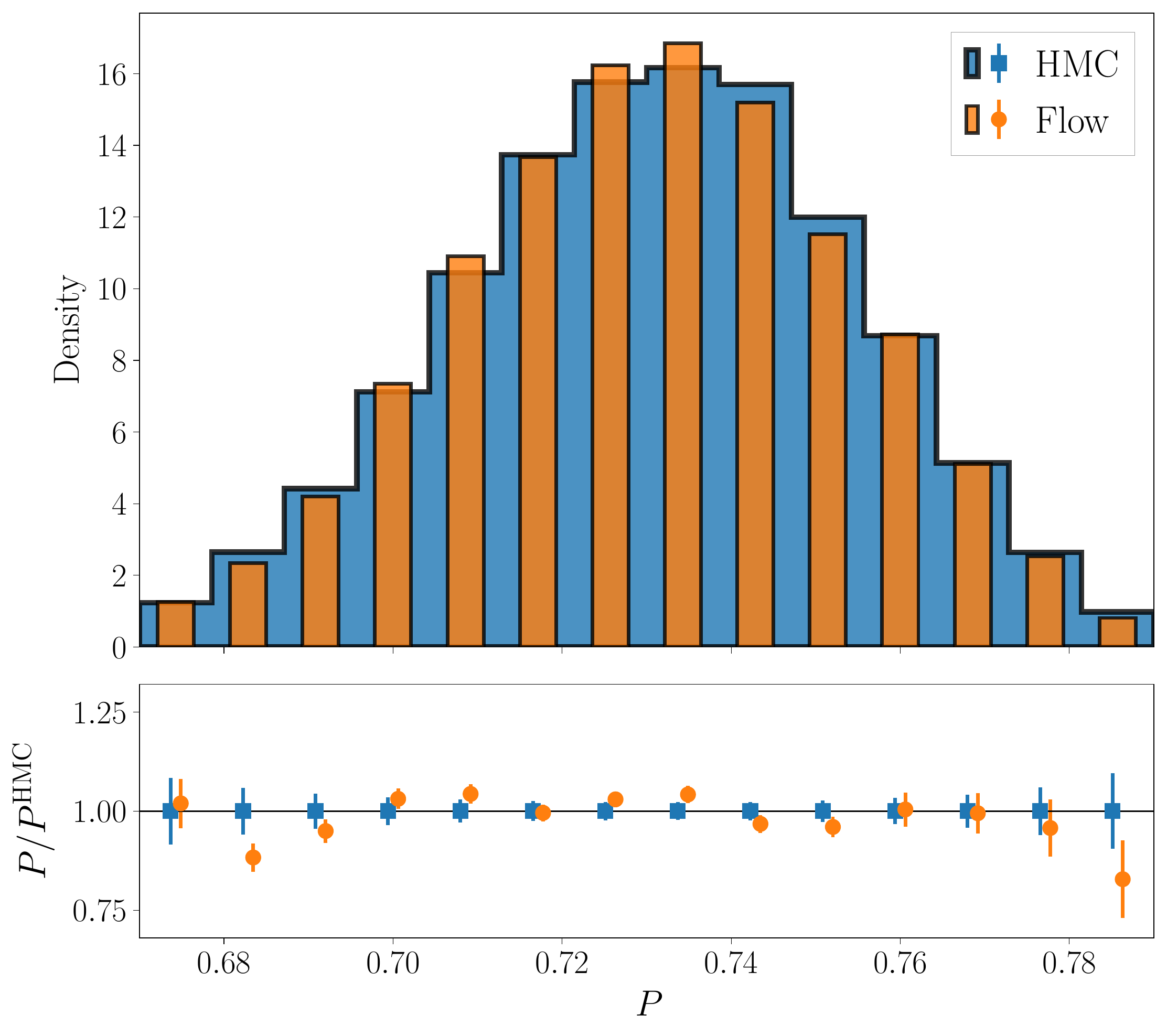}
    }
    \subfloat[Topological charge in U(1).]{%
     \includegraphics[width=0.48\textwidth]{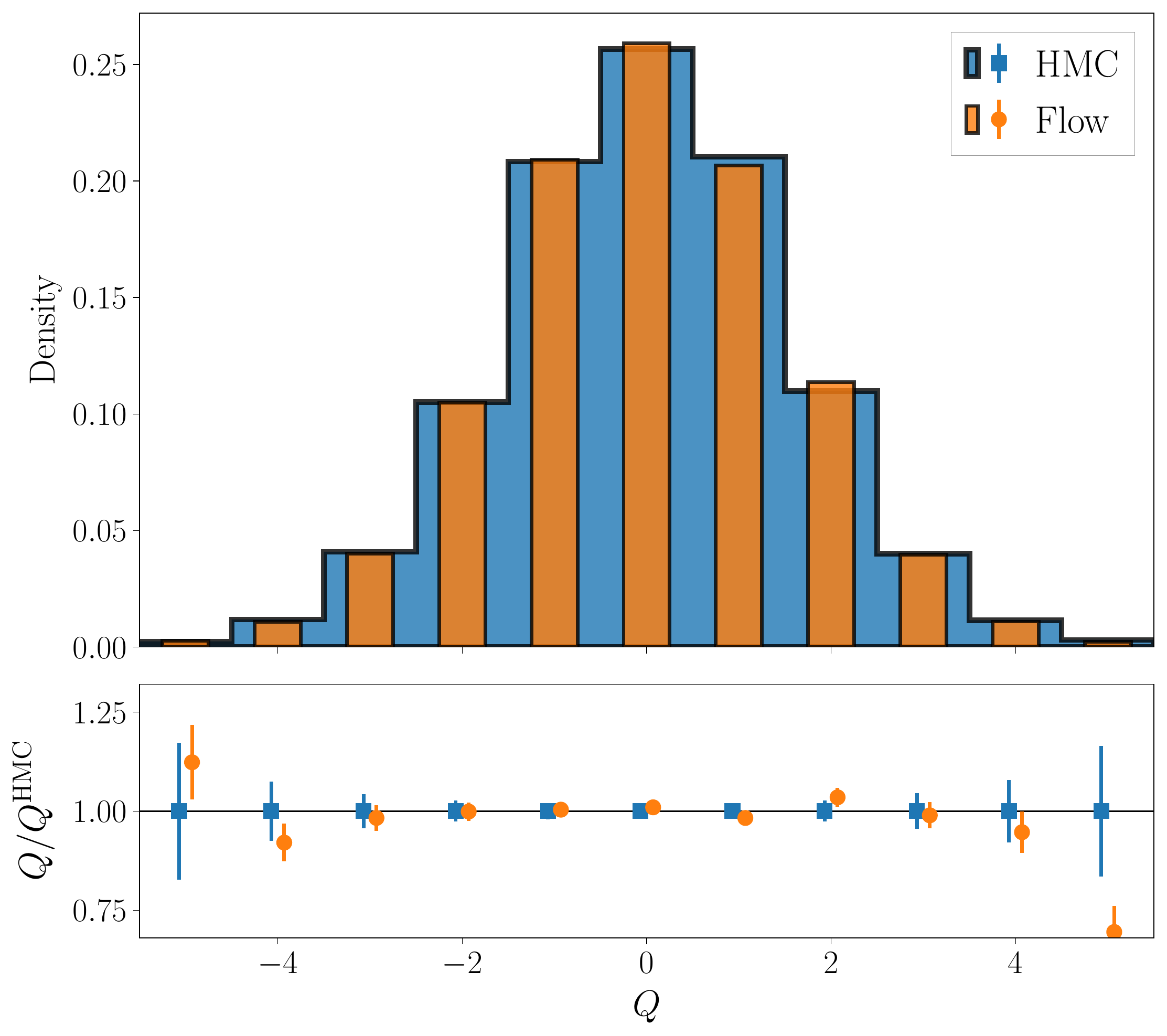}
    }
    \caption{ Comparison of distributions for two observables in the Schwinger model. The upper panels show density histograms comparing the distributions from HMC and flow-based sampling using reweighting. The lower panels show the ratio of the counts in each bin divided by the central value of the counts obtained in the same bin using the HMC ensemble. The uncertainties are estimated from 200 bootstrap ensembles.
    Histograms are constructed using comparable statistics, corresponding to approximately $\sim 10k$ independent configurations: 52k configurations from the flow model of \Cref{fig:ESS16} using $N_{\rm{pf}}=64$ pseudofermion samples with an $\text{ESS} \sim 30\%$, and 130k HMC configurations thinned by a factor of 10 to yield independent samples. }
    \label{fig:histsu1}
\end{figure*}

\begin{figure*}[t!]
     \centering
     \subfloat[Average plaquette in SU(3).]{%
     \includegraphics[width=0.48\textwidth]{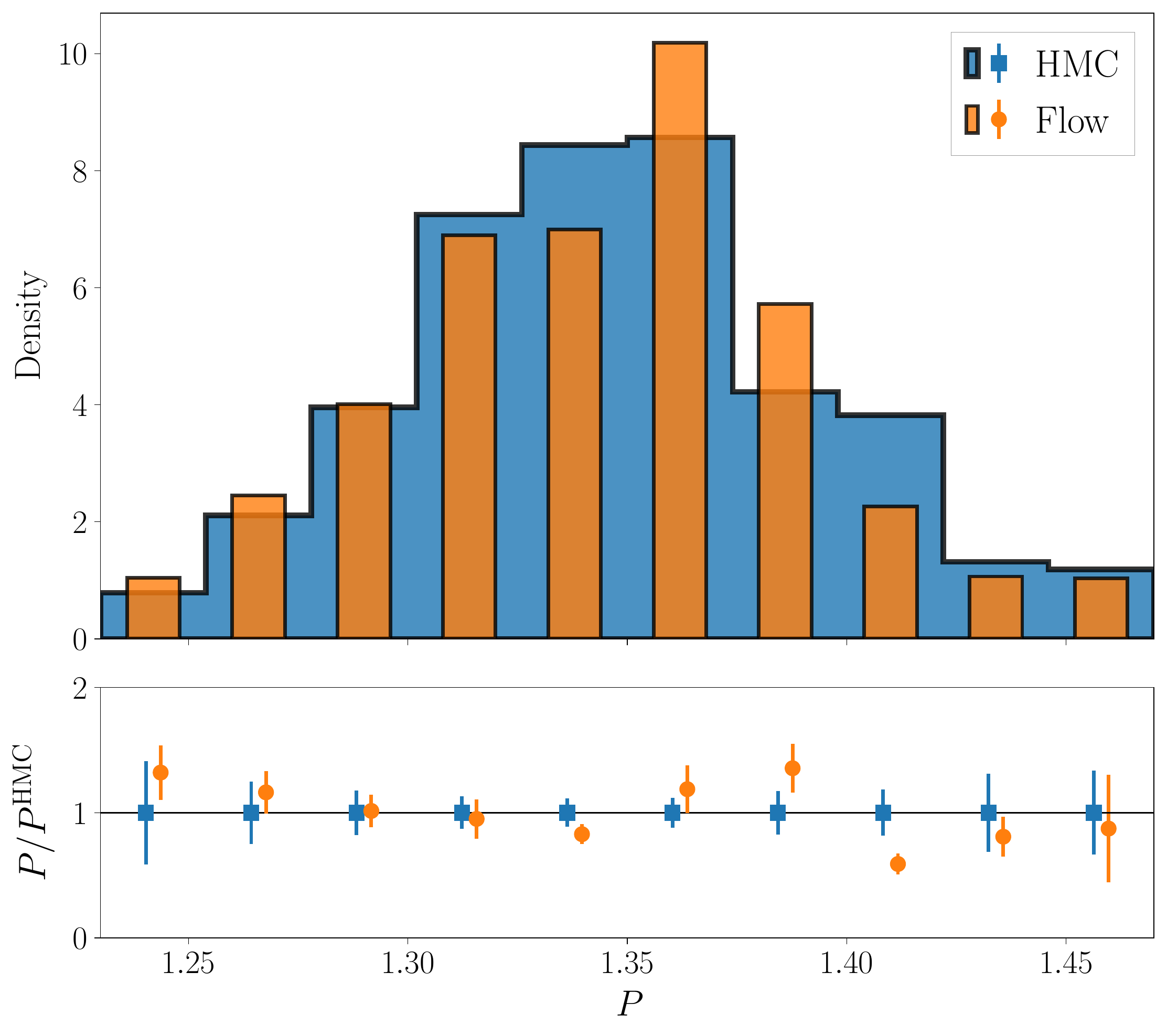}
    }
    \subfloat[Pseudoscalar correlator at $x_0=7$ in SU(3).]{%
     \includegraphics[width=0.48\textwidth]{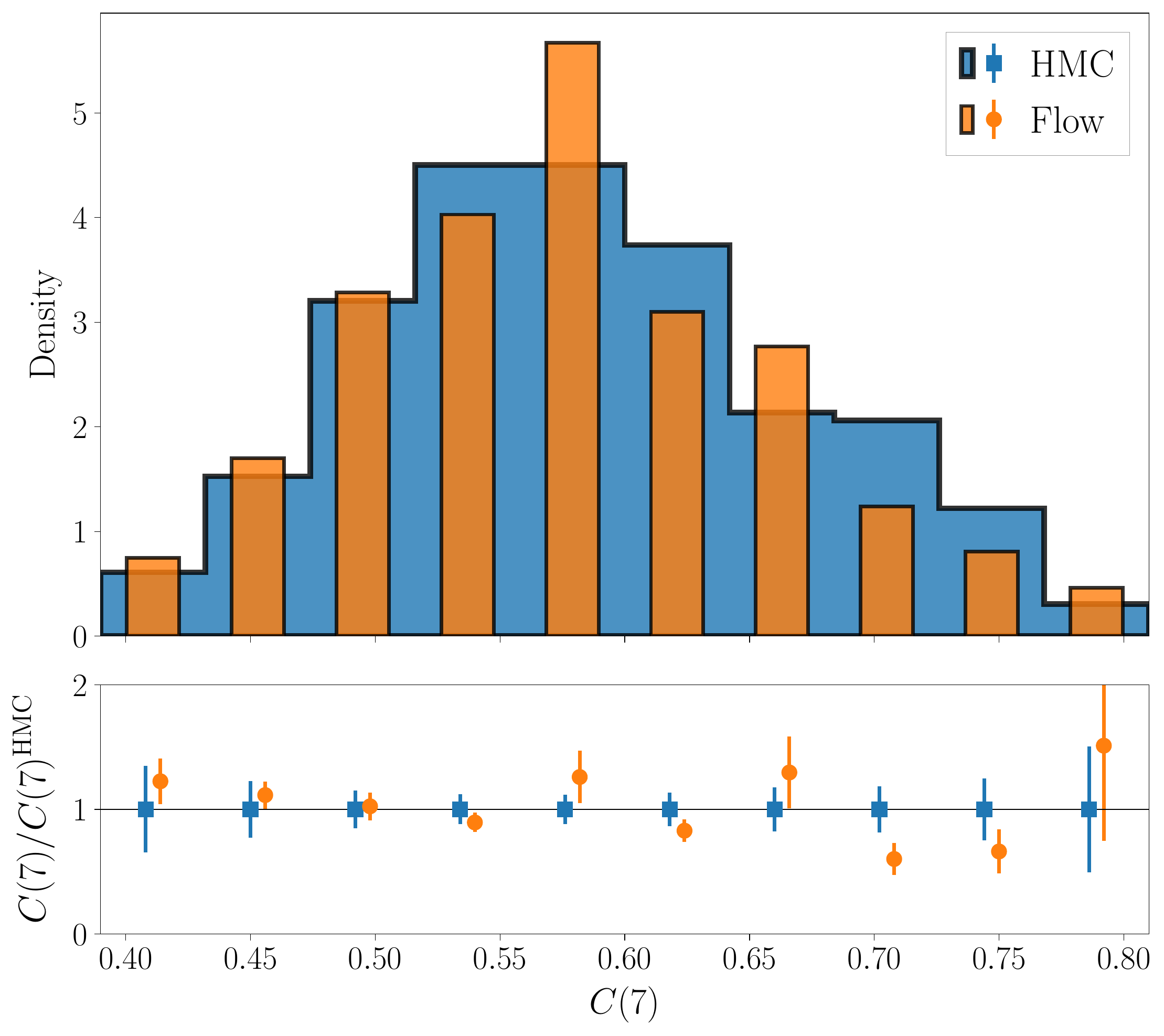}
    } \hfill
    \caption{ Same as \Cref{fig:histsu1} but for two observables for the 2D fermionic SU(3) theory. 
    Histograms are constructed using comparable statistics, corresponding to approximately $\sim 500$ independent configurations: 41k configurations from the flow model of \Cref{fig:ESS16} using a single pseudofermion sample with an $\text{ESS} \sim 3\%$, and 3200 HMC configurations thinned by a factor of 10 to yield independent samples. }
    \label{fig:histssu3}
\end{figure*}

In this appendix, we compare distributions of values for observables computed on gauge field configurations generated using HMC to the those computed on configurations from flow models (reweighted). \Cref{fig:histsu1,fig:histssu3} display histograms comparing two different observables for each of the two fermionic gauge theories studied in this work. In all cases, the observables computed on configurations obtained using both sampling algorithms are statistically consistent.

For the Schwinger model, the flow model of \Cref{fig:ESS16} is used. The observables compared are the average plaquette, 
    \begin{equation}
        P = \frac{1}{L^2} \sum_x \text{Re } \text{tr } P_{\mu\nu}(x) , \label{eq:avgplaq}
    \end{equation}
and the topological charge,
    \begin{equation}
        Q=\frac{1}{2 \pi} \sum_x \text{Im} \log P(x).
    \end{equation}
For the SU(3) gauge theory, the flow model of \Cref{fig:SU316} is used. As an example, two observables are compared: the average plaquette, as in \Cref{eq:avgplaq}, and the pseudoscalar correlator,
    \begin{equation}
        C\left(x_0\right)=-\sum_{x_1, y_1}\left\langle\left[\bar{u} \gamma_5 d\right]\left(x_0, x_1 \right)\left[\bar{d} \gamma_5 u\right](0, y_1)\right\rangle.
    \end{equation}


\bibliographystyle{utphys}
\bibliography{main}

\providecommand{\href}[2]{#2}\begingroup\raggedright\begin{thebibliography}{10}

\bibitem{Detmold:2019ghl}
{USQCD} Collaboration, W.~Detmold, R.~G. Edwards, J.~J. Dudek, M.~Engelhardt,
  H.-W. Lin, S.~Meinel, K.~Orginos, and P.~Shanahan
  \href{http://dx.doi.org/10.1140/epja/i2019-12902-4}{{\em Eur. Phys. J. A}
  {\bfseries 55} no.~11, (2019) 193},
  \href{http://arxiv.org/abs/1904.09512}{{\ttfamily arXiv:1904.09512
  [hep-lat]}}.

\bibitem{USQCD:2019hyg}
{USQCD} Collaboration, C.~Lehner {\em et~al.}
  \href{http://dx.doi.org/10.1140/epja/i2019-12891-2}{{\em Eur. Phys. J. A}
  {\bfseries 55} no.~11, (2019) 195},
  \href{http://arxiv.org/abs/1904.09479}{{\ttfamily arXiv:1904.09479
  [hep-lat]}}.

\bibitem{Kronfeld:2019nfb}
{USQCD} Collaboration, A.~S. Kronfeld, D.~G. Richards, W.~Detmold, R.~Gupta,
  H.-W. Lin, K.-F. Liu, A.~S. Meyer, R.~Sufian, and S.~Syritsyn
  \href{http://dx.doi.org/10.1140/epja/i2019-12916-x}{{\em Eur. Phys. J. A}
  {\bfseries 55} no.~11, (2019) 196},
  \href{http://arxiv.org/abs/1904.09931}{{\ttfamily arXiv:1904.09931
  [hep-lat]}}.

\bibitem{Boyle:2022uba}
P.~A. Boyle {\em et~al.} in {\em {2022 Snowmass Summer Study}}.
\newblock 5, 2022.
\newblock \href{http://arxiv.org/abs/2205.15373}{{\ttfamily arXiv:2205.15373
  [hep-lat]}}.

\bibitem{Joo:2019byq}
{USQCD} Collaboration, B.~Jo\'o, C.~Jung, N.~H. Christ, W.~Detmold, R.~Edwards,
  M.~Savage, and P.~Shanahan
  \href{http://dx.doi.org/10.1140/epja/i2019-12919-7}{{\em Eur. Phys. J. A}
  {\bfseries 55} no.~11, (2019) 199},
  \href{http://arxiv.org/abs/1904.09725}{{\ttfamily arXiv:1904.09725
  [hep-lat]}}.

\bibitem{Boyda:2022nmh}
D.~Boyda {\em et~al.} in {\em {2022 Snowmass Summer Study}}.
\newblock 2, 2022.
\newblock \href{http://arxiv.org/abs/2202.05838}{{\ttfamily arXiv:2202.05838
  [hep-lat]}}.

\bibitem{Wang2017}
L.~Wang \href{http://dx.doi.org/10.1103/PhysRevE.96.051301}{{\em Phys. Rev. E}
  {\bfseries 96} (Nov, 2017) 051301}.
  \url{https://link.aps.org/doi/10.1103/PhysRevE.96.051301}.

\bibitem{Huang:2017}
L.~Huang and L.~Wang \href{http://dx.doi.org/10.1103/physrevb.95.035105}{{\em
  Physical Review B} {\bfseries 95} no.~3, (Jan, 2017) --}.
  \url{http://dx.doi.org/10.1103/PhysRevB.95.035105}.

\bibitem{song2017nice}
J.~Song, S.~Zhao, and S.~Ermon in {\em Advances in Neural Information
  Processing Systems}, pp.~5140--5150.
\newblock 2017.

\bibitem{Tanaka:2017niz}
A.~Tanaka and A.~Tomiya \href{http://arxiv.org/abs/1712.03893}{{\ttfamily
  arXiv:1712.03893 [hep-lat]}}.

\bibitem{levy2018generalizing}
D.~Levy, M.~D. Hoffman, and J.~Sohl-Dickstein
  \href{http://arxiv.org/abs/1711.09268}{{\ttfamily arXiv:1711.09268
  [stat.ML]}}.

\bibitem{Pawlowski:2018qxs}
J.~M. Pawlowski and J.~M. Urban
  \href{http://dx.doi.org/10.1088/2632-2153/abae73}{{\em Mach. Learn. Sci.
  Tech.} {\bfseries 1} (2020) 045011},
  \href{http://arxiv.org/abs/1811.03533}{{\ttfamily arXiv:1811.03533
  [hep-lat]}}.

\bibitem{Cossu:2018pxj}
G.~Cossu, L.~Del~Debbio, T.~Giani, A.~Khamseh, and M.~Wilson
  \href{http://dx.doi.org/10.1103/PhysRevB.100.064304}{{\em Phys. Rev. B}
  {\bfseries 100} no.~6, (2019) 064304},
  \href{http://arxiv.org/abs/1810.11503}{{\ttfamily arXiv:1810.11503
  [physics.comp-ph]}}.

\bibitem{Wu:2019}
D.~Wu, L.~Wang, and P.~Zhang
  \href{http://dx.doi.org/10.1103/PhysRevLett.122.080602}{{\em Phys. Rev.
  Lett.} {\bfseries 122} (Feb, 2019) 080602}.
  \url{https://link.aps.org/doi/10.1103/PhysRevLett.122.080602}.

\bibitem{Bachtis:2020dmf}
D.~Bachtis, G.~Aarts, and B.~Lucini
  \href{http://dx.doi.org/10.1103/PhysRevE.102.033303}{{\em Phys. Rev. E}
  {\bfseries 102} no.~3, (2020) 033303},
  \href{http://arxiv.org/abs/2004.14341}{{\ttfamily arXiv:2004.14341
  [cond-mat.stat-mech]}}.

\bibitem{Nagai:2020jar}
Y.~Nagai, A.~Tanaka, and A.~Tomiya
  \href{http://arxiv.org/abs/2010.11900}{{\ttfamily arXiv:2010.11900
  [hep-lat]}}.

\bibitem{Tomiya:2021ywc}
A.~Tomiya and Y.~Nagai \href{http://arxiv.org/abs/2103.11965}{{\ttfamily
  arXiv:2103.11965 [hep-lat]}}.

\bibitem{Bachtis:2021eww}
D.~Bachtis, G.~Aarts, F.~Di~Renzo, and B.~Lucini {\em Phys. Rev. Lett. (to
  appear)} (7, 2021) , \href{http://arxiv.org/abs/2107.00466}{{\ttfamily
  arXiv:2107.00466 [hep-lat]}}.

\bibitem{Wu:2021tfb}
D.~Wu, R.~Rossi, and G.~Carleo
  \href{http://arxiv.org/abs/2105.05650}{{\ttfamily arXiv:2105.05650
  [cond-mat.stat-mech]}}.

\bibitem{rezende2016variational}
D.~J. Rezende and S.~Mohamed \href{http://arxiv.org/abs/1505.05770}{{\ttfamily
  arXiv:1505.05770 [stat.ML]}}.

\bibitem{dinh2017density}
L.~Dinh, J.~Sohl-Dickstein, and S.~Bengio
  \href{http://arxiv.org/abs/1605.08803}{{\ttfamily arXiv:1605.08803 [cs.LG]}}.

\bibitem{JMLR:v22:19-1028}
G.~Papamakarios, E.~Nalisnick, D.~J. Rezende, S.~Mohamed, and
  B.~Lakshminarayanan {\em Journal of Machine Learning Research} {\bfseries 22}
  no.~57, (2021) 1--64.

\bibitem{LiWang2018NNRG}
S.-H. Li and L.~Wang
  \href{http://dx.doi.org/10.1103/PhysRevLett.121.260601}{{\em Phys. Rev.
  Lett.} {\bfseries 121} (Dec, 2018) 260601}.
  \url{https://link.aps.org/doi/10.1103/PhysRevLett.121.260601}.

\bibitem{Albergo:2019eim}
M.~S. Albergo, G.~Kanwar, and P.~E. Shanahan
  \href{http://dx.doi.org/10.1103/PhysRevD.100.034515}{{\em Phys. Rev. D}
  {\bfseries 100} no.~3, (2019) 034515},
  \href{http://arxiv.org/abs/1904.12072}{{\ttfamily arXiv:1904.12072
  [hep-lat]}}.

\bibitem{Kanwar:2020xzo}
G.~Kanwar, M.~S. Albergo, D.~Boyda, K.~Cranmer, D.~C. Hackett, S.~Racani\`ere,
  D.~J. Rezende, and P.~E. Shanahan
  \href{http://dx.doi.org/10.1103/PhysRevLett.125.121601}{{\em Phys. Rev.
  Lett.} {\bfseries 125} no.~12, (2020) 121601},
  \href{http://arxiv.org/abs/2003.06413}{{\ttfamily arXiv:2003.06413
  [hep-lat]}}.

\bibitem{Boyda:2020hsi}
D.~Boyda, G.~Kanwar, S.~Racani\`ere, D.~J. Rezende, M.~S. Albergo, K.~Cranmer,
  D.~C. Hackett, and P.~E. Shanahan
  \href{http://dx.doi.org/10.1103/PhysRevD.103.074504}{{\em Phys. Rev. D}
  {\bfseries 103} no.~7, (2021) 074504},
  \href{http://arxiv.org/abs/2008.05456}{{\ttfamily arXiv:2008.05456
  [hep-lat]}}.

\bibitem{Hackett:2021idh}
D.~C. Hackett, C.-C. Hsieh, M.~S. Albergo, D.~Boyda, J.-W. Chen, K.-F. Chen,
  K.~Cranmer, G.~Kanwar, and P.~E. Shanahan
  \href{http://arxiv.org/abs/2107.00734}{{\ttfamily arXiv:2107.00734
  [hep-lat]}}.

\bibitem{Albergo:2021bna}
M.~S. Albergo, G.~Kanwar, S.~Racani\`ere, D.~J. Rezende, J.~M. Urban, D.~Boyda,
  K.~Cranmer, D.~C. Hackett, and P.~E. Shanahan
  \href{http://dx.doi.org/10.1103/PhysRevD.104.114507}{{\em Phys. Rev. D}
  {\bfseries 104} no.~11, (2021) 114507},
  \href{http://arxiv.org/abs/2106.05934}{{\ttfamily arXiv:2106.05934
  [hep-lat]}}.

\bibitem{Albergo:2021vyo}
M.~S. Albergo, D.~Boyda, D.~C. Hackett, G.~Kanwar, K.~Cranmer, S.~Racani\`ere,
  D.~J. Rezende, and P.~E. Shanahan
  \href{http://arxiv.org/abs/2101.08176}{{\ttfamily arXiv:2101.08176
  [hep-lat]}}.

\bibitem{Albergo:2022qfi}
M.~S. Albergo, D.~Boyda, K.~Cranmer, D.~C. Hackett, G.~Kanwar, S.~Racani\`ere,
  D.~J. Rezende, F.~Romero-L\'opez, P.~E. Shanahan, and J.~M. Urban
  \href{http://arxiv.org/abs/2202.11712}{{\ttfamily arXiv:2202.11712
  [hep-lat]}}.

\bibitem{Nicoli:2020evf}
K.~A. Nicoli, S.~Nakajima, N.~Strodthoff, W.~Samek, K.-R. M\"uller, and
  P.~Kessel \href{http://dx.doi.org/10.1103/PhysRevE.101.023304}{{\em Phys.
  Rev. E} {\bfseries 101} no.~2, (2020) 023304},
  \href{http://arxiv.org/abs/1910.13496}{{\ttfamily arXiv:1910.13496
  [cond-mat.stat-mech]}}.

\bibitem{Nicoli:2020njz}
K.~A. Nicoli, C.~J. Anders, L.~Funcke, T.~Hartung, K.~Jansen, P.~Kessel,
  S.~Nakajima, and P.~Stornati
  \href{http://arxiv.org/abs/2007.07115}{{\ttfamily arXiv:2007.07115
  [hep-lat]}}.

\bibitem{Foreman:2021ixr}
S.~Foreman, X.-Y. Jin, and J.~C. Osborn in {\em {9th International Conference
  on Learning Representations}}.
\newblock May, 2021.
\newblock \href{http://arxiv.org/abs/2105.03418}{{\ttfamily arXiv:2105.03418
  [hep-lat]}}.

\bibitem{Foreman:2021ljl}
S.~Foreman, T.~Izubuchi, L.~Jin, X.-Y. Jin, J.~C. Osborn, and A.~Tomiya in {\em
  {38th International Symposium on Lattice Field Theory}}.
\newblock Dec, 2021.
\newblock \href{http://arxiv.org/abs/2112.01586}{{\ttfamily arXiv:2112.01586
  [cs.LG]}}.

\bibitem{Foreman:2021rhs}
S.~Foreman, X.-Y. Jin, and J.~C. Osborn in {\em {38th International Symposium
  on Lattice Field Theory}}.
\newblock 12, 2021.
\newblock \href{http://arxiv.org/abs/2112.01582}{{\ttfamily arXiv:2112.01582
  [hep-lat]}}.

\bibitem{DelDebbio:2021qwf}
L.~Del~Debbio, J.~M. Rossney, and M.~Wilson
  \href{http://arxiv.org/abs/2105.12481}{{\ttfamily arXiv:2105.12481
  [hep-lat]}}.

\bibitem{Gabrie:2021tlu}
M.~Gabri\'e, G.~M. Rotskoff, and E.~Vanden-Eijnden
  \href{http://arxiv.org/abs/2105.12603}{{\ttfamily arXiv:2105.12603
  [physics.data-an]}}.

\bibitem{deHaan:2021erb}
P.~de~Haan, C.~Rainone, M.~C.~N. Cheng, and R.~Bondesan
  \href{http://arxiv.org/abs/2110.02673}{{\ttfamily arXiv:2110.02673 [cs.LG]}}.

\bibitem{Lawrence:2021izu}
S.~Lawrence and Y.~Yamauchi
  \href{http://dx.doi.org/10.1103/PhysRevD.103.114509}{{\em Phys. Rev. D}
  {\bfseries 103} no.~11, (2021) 114509},
  \href{http://arxiv.org/abs/2101.05755}{{\ttfamily arXiv:2101.05755
  [hep-lat]}}.

\bibitem{Jin:2022bgq}
X.-Y. Jin in {\em {38th International Symposium on Lattice Field Theory}}.
\newblock 1, 2022.
\newblock \href{http://arxiv.org/abs/2201.01862}{{\ttfamily arXiv:2201.01862
  [hep-lat]}}.

\bibitem{Pawlowski:2022rdn}
J.~M. Pawlowski and J.~M. Urban
  \href{http://arxiv.org/abs/2203.01243}{{\ttfamily arXiv:2203.01243
  [hep-lat]}}.

\bibitem{Finkenrath:2022ogg}
J.~Finkenrath \href{http://arxiv.org/abs/2201.02216}{{\ttfamily
  arXiv:2201.02216 [hep-lat]}}.

\bibitem{Gerdes:2022eve}
M.~Gerdes, P.~de~Haan, C.~Rainone, R.~Bondesan, and M.~C.~N. Cheng
  \href{http://arxiv.org/abs/2207.00283}{{\ttfamily arXiv:2207.00283
  [hep-lat]}}.

\bibitem{Singha:2022lpi}
A.~Singha, D.~Chakrabarti, and V.~Arora
  \href{http://arxiv.org/abs/2207.00980}{{\ttfamily arXiv:2207.00980
  [hep-lat]}}.

\bibitem{Matthews:2022sds}
A.~G. D.~G. Matthews, M.~Arbel, D.~J. Rezende, and A.~Doucet
  \href{http://arxiv.org/abs/2201.13117}{{\ttfamily arXiv:2201.13117
  [stat.ML]}}.

\bibitem{Caselle:2022acb}
M.~Caselle, E.~Cellini, A.~Nada, and M.~Panero
  \href{http://arxiv.org/abs/2201.08862}{{\ttfamily arXiv:2201.08862
  [hep-lat]}}.

\bibitem{favoni2020lattice}
M.~Favoni, A.~Ipp, D.~I. M{\"u}ller, and D.~Schuh {\em arXiv preprint
  arXiv:2012.12901} (2020) .

\bibitem{Wilson:1974sk}
K.~G. Wilson \href{http://dx.doi.org/10.1103/PhysRevD.10.2445}{{\em Phys. Rev.
  D} {\bfseries 10} (1974) 2445--2459}.

\bibitem{Wilson:1975id}
K.~G. Wilson in {\em {13th International School of Subnuclear Physics: New
  Phenomena in Subnuclear Physics}}.
\newblock Nov, 1975.

\bibitem{10.5555/865018}
J.~R. Shewchuk tech. rep., USA, 1994.

\bibitem{Kullback:1951}
S.~Kullback and R.~A. Leibler
  \href{http://dx.doi.org/10.1214/aoms/1177729694}{{\em The Annals of
  Mathematical Statistics} {\bfseries 22} no.~1, (1951) 79 -- 86}.

\bibitem{sabour2017dynamic}
S.~Sabour, N.~Frosst, and G.~E. Hinton {\em Advances in neural information
  processing systems} {\bfseries 30} (2017) .

\bibitem{DeGrand:1990dk}
T.~A. DeGrand and P.~Rossi
  \href{http://dx.doi.org/10.1016/0010-4655(90)90006-M}{{\em Comput. Phys.
  Commun.} {\bfseries 60} (1990) 211--214}.

\bibitem{Hasenbusch:2001ne}
M.~Hasenbusch \href{http://dx.doi.org/10.1016/S0370-2693(01)01102-9}{{\em Phys.
  Lett. B} {\bfseries 519} (2001) 177--182},
  \href{http://arxiv.org/abs/hep-lat/0107019}{{\ttfamily
  arXiv:hep-lat/0107019}}.

\bibitem{Andrieu:2009}
C.~Andrieu and G.~O. Roberts \href{http://dx.doi.org/10.1214/07-AOS574}{{\em
  The Annals of Statistics} {\bfseries 37} no.~2, (2009) 697 -- 725},
  \href{http://arxiv.org/abs/0903.5480}{{\ttfamily arXiv:0903.5480 [math.ST]}}.

\bibitem{Schwinger:1962tp}
J.~S. Schwinger \href{http://dx.doi.org/10.1103/PhysRev.128.2425}{{\em Phys.
  Rev.} {\bfseries 128} (1962) 2425--2429}.

\bibitem{Coleman:1975pw}
S.~R. Coleman, R.~Jackiw, and L.~Susskind
  \href{http://dx.doi.org/10.1016/0003-4916(75)90212-2}{{\em Annals Phys.}
  {\bfseries 93} (1975) 267}.

\bibitem{Smit:1987fq}
J.~Smit and J.~C. Vink
  \href{http://dx.doi.org/10.1016/0550-3213(88)90215-5}{{\em Nucl. Phys. B}
  {\bfseries 303} (1988) 36--56}.

\bibitem{Dilger:1992yn}
H.~Dilger \href{http://dx.doi.org/10.1016/0370-2693(92)90692-W}{{\em Phys.
  Lett. B} {\bfseries 294} (1992) 263--268}.

\bibitem{Dilger:1994ma}
H.~Dilger \href{http://dx.doi.org/10.1142/S0129183195000101}{{\em Int. J. Mod.
  Phys. C} {\bfseries 6} (1995) 123--134},
  \href{http://arxiv.org/abs/hep-lat/9408017}{{\ttfamily
  arXiv:hep-lat/9408017}}.

\bibitem{Durr:2012te}
S.~Durr \href{http://dx.doi.org/10.1103/PhysRevD.85.114503}{{\em Phys. Rev. D}
  {\bfseries 85} (2012) 114503},
  \href{http://arxiv.org/abs/1203.2560}{{\ttfamily arXiv:1203.2560 [hep-lat]}}.

\bibitem{Funcke:2019zna}
L.~Funcke, K.~Jansen, and S.~K\"uhn
  \href{http://dx.doi.org/10.1103/PhysRevD.101.054507}{{\em Phys. Rev. D}
  {\bfseries 101} no.~5, (2020) 054507},
  \href{http://arxiv.org/abs/1908.00551}{{\ttfamily arXiv:1908.00551
  [hep-lat]}}.

\bibitem{Butt:2019uul}
N.~Butt, S.~Catterall, Y.~Meurice, R.~Sakai, and J.~Unmuth-Yockey
  \href{http://dx.doi.org/10.1103/PhysRevD.101.094509}{{\em Phys. Rev. D}
  {\bfseries 101} no.~9, (2020) 094509},
  \href{http://arxiv.org/abs/1911.01285}{{\ttfamily arXiv:1911.01285
  [hep-lat]}}.

\bibitem{Banuls:2019bmf}
M.~C. Ba\~nuls {\em et~al.}
  \href{http://dx.doi.org/10.1140/epjd/e2020-100571-8}{{\em Eur. Phys. J. D}
  {\bfseries 74} no.~8, (2020) 165},
  \href{http://arxiv.org/abs/1911.00003}{{\ttfamily arXiv:1911.00003
  [quant-ph]}}.

\bibitem{Albandea:2021lvl}
D.~Albandea, P.~Hern\'andez, A.~Ramos, and F.~Romero-L\'opez
  \href{http://dx.doi.org/10.1140/epjc/s10052-021-09677-6}{{\em Eur. Phys. J.
  C} {\bfseries 81} no.~10, (2021) 873},
  \href{http://arxiv.org/abs/2106.14234}{{\ttfamily arXiv:2106.14234
  [hep-lat]}}.

\bibitem{Hartung:2021wqg}
T.~Hartung, K.~Jansen, F.~Y. Kuo, H.~Le\"ovey, D.~Nuyens, and I.~H. Sloan
  \href{http://arxiv.org/abs/2112.05069}{{\ttfamily arXiv:2112.05069
  [hep-lat]}}.

\bibitem{Eichhorn:2021ccz}
T.~Eichhorn and C.~Hoelbling in {\em {38th International Symposium on Lattice
  Field Theory}}.
\newblock Dec, 2021.
\newblock \href{http://arxiv.org/abs/2112.05188}{{\ttfamily arXiv:2112.05188
  [hep-lat]}}.

\bibitem{Abbott:2022hkm}
R.~Abbott {\em et~al.} in {\em {39th International Symposium on Lattice Field
  Theory}}.
\newblock 8, 2022.
\newblock \href{http://arxiv.org/abs/2208.03832}{{\ttfamily arXiv:2208.03832
  [hep-lat]}}.

\bibitem{scaling}
R.~Abbott {\em et~al.} {\em Aspects of scaling and scalability for flow-based
  sampling of lattice QCD (in preparation)} (2022) .

\bibitem{reuther2018interactive}
A.~Reuther, J.~Kepner, C.~Byun, S.~Samsi, W.~Arcand, D.~Bestor, B.~Bergeron,
  V.~Gadepally, M.~Houle, M.~Hubbell, {\em et~al.}
  \href{http://dx.doi.org/10.1109/hpec.2018.8547629}{{\em 2018 IEEE High
  Performance extreme Computing Conference (HPEC)} (Sep, 2018) 1--6},
  \href{http://arxiv.org/abs/1807.07814}{{\ttfamily arXiv:1807.07814 [cs.DC]}}.

\bibitem{NEURIPS2019_9015}
A.~Paszke {\em et~al.} in {\em Advances in Neural Information Processing
  Systems 32}, H.~Wallach, H.~Larochelle, A.~Beygelzimer, F.~d\textquotesingle
  Alch\'{e}-Buc, E.~Fox, and R.~Garnett, eds., pp.~8024--8035.
\newblock Curran Associates, Inc., 2019.
\newblock
  \url{http://papers.neurips.cc/paper/9015-pytorch-an-imperative-style-high-performance\\-deep-learning-library.pdf}.

\bibitem{jax2018github}
J.~Bradbury, R.~Frostig, P.~Hawkins, M.~J. Johnson, C.~Leary, D.~Maclaurin,
  G.~Necula, A.~Paszke, J.~Vander{P}las, S.~Wanderman-{M}ilne, and Q.~Zhang,
  2018.
\newblock \url{http://github.com/google/jax}.

\bibitem{haiku2020github}
T.~Hennigan, T.~Cai, T.~Norman, and I.~Babuschkin, 2020.
\newblock \url{http://github.com/deepmind/dm-haiku}.

\bibitem{sergeev2018horovod}
A.~{Sergeev} and M.~{Del Balso}
  \href{http://arxiv.org/abs/1802.05799}{{\ttfamily arXiv:1802.05799 [cs.LG]}}.

\bibitem{harris2020array}
C.~R. Harris, K.~J. Millman, S.~J. Van Der~Walt, R.~Gommers, P.~Virtanen,
  D.~Cournapeau, E.~Wieser, J.~Taylor, S.~Berg, N.~J. Smith, {\em et~al.} {\em
  Nature} {\bfseries 585} no.~7825, (2020) 357--362.

\bibitem{2020SciPy-NMeth}
P.~Virtanen, R.~Gommers, T.~E. Oliphant, M.~Haberland, T.~Reddy, D.~Cournapeau,
  E.~Burovski, P.~Peterson, W.~Weckesser, J.~Bright, {\em et~al.} {\em Nature
  methods} {\bfseries 17} no.~3, (2020) 261--272.

\bibitem{Hunter:2007}
J.~D. Hunter \href{http://dx.doi.org/10.1109/MCSE.2007.55}{{\em Computing in
  Science \& Engineering} {\bfseries 9} no.~3, (2007) 90--95}.

\bibitem{loshchilov2017decoupled}
I.~Loshchilov and F.~Hutter {\em arXiv preprint arXiv:1711.05101} (2017) .

\bibitem{2015arXiv151107289C}
D.-A. {Clevert}, T.~{Unterthiner}, and S.~{Hochreiter} {\em arXiv e-prints}
  (Nov., 2015) arXiv:1511.07289,
  \href{http://arxiv.org/abs/1511.07289}{{\ttfamily arXiv:1511.07289 [cs.LG]}}.

\end{thebibliography}\endgroup

\end{document}